\documentclass[aps,prl,twocolumn,preprintnumbers,amsmath,amssymb,superscriptaddress]{revtex4-1}
\usepackage[english]{babel}
\usepackage{blindtext}
\usepackage[utf8]{inputenc}
\usepackage{graphicx}
\usepackage{bm}
\usepackage{latexsym}
\usepackage{soul}
\usepackage{xcolor}
\usepackage{physics}
\usepackage{amsfonts}
\usepackage{bbold}
\usepackage{scalerel}
\usepackage{appendix}
\usepackage{makecell}
\usepackage[normalem]{ulem}
\usepackage[colorlinks = true,
            linkcolor = blue,
            urlcolor  = blue,
            citecolor = blue,
            anchorcolor = blue]{hyperref}

\usepackage{pdfpages}
\usepackage{etoolbox} 

\makeatletter
\patchcmd{\@outputpage@head}{\@ifx{\LS@rot\@undefined}{}{\LS@rot}}{}{}{}
\makeatother

\usepackage{tikz}
\graphicspath{{figures_PRL/}}
\newcommand{\bk}{\mathbf{k}}

\newcommand{\bq}{\mathbf{q}}
\newcommand{\bQ}{\mathbf{Q}}
\newcommand{\up}{\uparrow}
\newcommand{\down}{\downarrow}

\definecolor{green(html/cssgreen)}{rgb}{0.0, 0.5, 0.0}

\allowdisplaybreaks

\usepackage{xr}
\makeatletter
\newcommand*{\addFileDependency}[1]{
 \typeout{(#1)}
 \@addtofilelist{#1}
  \IfFileExists{#1}{}{\typeout{No file #1.}}
}
\makeatother

\newcommand*{\myexternaldocument}[1]{
    \externaldocument{#1}
    \addFileDependency{#1.tex}
    \addFileDependency{#1.aux}
}
\myexternaldocument{methods}
\begin{document}

\title{Competing phases in kagome magnet FeGe from functional renormalization}%

\date{\today}

\author{Pietro M. Bonetti}
\thanks{}
\affiliation{Department of Physics, Harvard University, Cambridge MA 02138, USA}
\affiliation{Max Planck Institute for Solid State Research, Heisenbergstrasse 1, D-70569 Stuttgart, Germany}

\author{Yi Jiang}
\thanks{}
\affiliation{Donostia International Physics Center (DIPC), Paseo Manuel de Lardizabal. 20018, San Sebastian, Spain}

\author{Haoyu Hu}
\thanks{}
\affiliation{Department of Physics, Princeton University, Princeton, NJ 08544, USA}

\author{Dumitru C\u alug\u aru}
\thanks{}
\affiliation{Department of Physics, Princeton University, Princeton, NJ 08544, USA}

\author{Michael M. Scherer}
\thanks{}
\affiliation{Theoretische Physik III, Ruhr-Universit\"at Bochum, D-44801 Bochum, Germany}

\author{B. Andrei Bernevig}
\thanks{}
\affiliation{Department of Physics, Princeton University, Princeton, NJ 08544, USA}
\affiliation{Donostia International Physics Center (DIPC), Paseo Manuel de Lardizabal. 20018, San Sebastian, Spain}
\affiliation{IKERBASQUE, Basque Foundation for Science, 48013 Bilbao, Spain}

\author{Laura Classen}
\thanks{}
\affiliation{Max Planck Institute for Solid State Research, Heisenbergstrasse 1, D-70569 Stuttgart, Germany}
\affiliation{Department of Physics, Technical University of Munich, D-85748 Garching, Germany}

\begin{abstract}
The discovery of a charge density wave in FeGe extends the discussion of the nature of charge order in kagome metals to a magnetic compound. Motivated by this observation, we combine density functional theory (DFT) and functional-renormalization-group calculations to study interaction-induced Fermi-surface instabilities of the magnetic state of FeGe. We argue that the leading intra-band contribution to electronic correlations are approximately 2D and come from Van Hove points at the projected $M$~points. By varying  parameters around DFT values, we determine a phase diagram for the quasi-2D scenario as function of on-site and nearest-neighbor interactions. We discuss universal aspects in the electronic mechanisms for the resulting phases, as well as the role of SU(2) symmetry breaking. We find FeGe to be in a regime of strong competition between $p$-wave charge density wave, $f$-wave pairing, and $d$-wave spin Pomeranchuk instabilities. This interplay can be influenced in favor of superconducting pairing for slightly increased nearest-neighbor interaction, suggesting a potential to induce superconductivity in FeGe. 
\end{abstract}

\maketitle

\paragraph{Introduction.}~Kagome metals constitute an important platform for studying electronic correlations. In recent years a rich phase structure was revealed in different such compounds
~\cite{neupert2022,Jiangreview2022,Hu2023,Wang_2024}. 
It is based on an interplay of different degrees of freedom 
which continues to challenge our understanding of the emergent many-body behavior, with multiple possible competing phases. In particular, the type of charge orders and the mechanism behind their formation is being intensely discussed, as well as their connection to magnetic or superconducting phases.  

In the prominent superconducting kagome family AV${}_3$Sb${}_5$ (A=K, Rb, Cs)~\cite{PhysRevMaterials.3.094407}, an exotic charge density wave (CDW) arises with wave vector connecting electronic Van Hove points (VHPs)
~\cite{Jiang2021,PhysRevX.11.031026,PhysRevResearch.4.023244,gupta2022types,Zhao2021,Kang2022,Xu2022,Kang2023,Hu2022,PhysRevB.106.L241106,
PhysRevResearch.4.033072,Xing2024optic}. It precedes a superconducting transition, while long-range magnetic order remains absent~\cite{PhysRevMaterials.5.034801,PhysRevLett.125.247002,Chen2021,Gupta2022,PhysRevLett.126.247001,Zhong2023phonon,Zhong2023nodeless}.  
An unconventional CDW was also discovered in the magnetic kagome metal FeGe~\cite{Teng2022,Teng2023,PhysRevResearch.6.013276,PhysRevLett.132.266505,yi2024polarizedcharge,shi2024,oh2024tunability,PhysRevLett.133.046502,han2024,subires2024}, which shows both similarities and differences to AV${}_3$Sb${}_5$. Remarkably, in contrast to AV${}_3$Sb${}_5$, the CDW in FeGe arises out of a collinear layer antiferromagnet (AFM) which has a (anti-)ferromagnetic alignment (between) within Fe kagome layers. 
Furthermore, the CDW precedes an incommensurate density wave with canted spins~\cite{doi:10.1143/JPSJ.18.589,Beckman_1972,Bernhard_1984,Bernhard_1988}. Yet, similar to AV${}_3$Sb${}_5$, the CDW vector connects VHPs, 
which, however, are now placed near the Fermi level only after the nearly flat bands of FeGe are split in the AFM phase~\cite{Teng2022,Teng2023,shi2024}. Simultaneously, a structural transition 
occurs and magnetic moments are enhanced~\cite{yi2024polarizedcharge,PhysRevMaterials.8.L080601}. Generally, experimental observations demonstrate that FeGe realizes a strong coupling of spin, charge, orbital, and lattice degrees of freedom in a single system~\cite{Miao2023,PhysRevLett.133.046502,PhysRevResearch.6.033222,PhysRevResearch.6.013276,PhysRevLett.132.256501,PhysRevLett.132.266505,shi2024,han2024,subires2024,oh2024tunability,yi2024polarizedcharge,PhysRevMaterials.8.L080601}. Naturally, the primary order parameter of the CDW remains debated, with no softening of phonons in the FeGe observed\cite{subires2024frustrated, korshunov2024pressure} or computed.

Here, we perform functional renormalization group (FRG) calculations for a \textit{ab-initio} band structure of FeGe to elucidate the role of electronic contributions in the CDW formation. The advantage of the FRG approach is an unbiased description of electronic charge, spin, and pairing correlations taking their mutual feedback into account. We do not include electron-phonon coupling. FRG was successfully applied to the single-orbital kagome lattice, pointing out the key role of sublattice interference for charge ordering~\cite{PhysRevLett.110.126405,PhysRevB.87.115135,profe2024kagome,schwemmer2024,PhysRevB.86.121105}. 
Different charge orders were also studied via Hartree-Fock 
\cite{PhysRevB.110.L041121}, or via phenomenological 
Landau analyses, sometimes 
combined with DFT for AV${}_3$Sb${}_5$~\cite{PhysRevB.106.144504,PhysRevB.104.214513,PhysRevB.107.205131,PhysRevLett.127.217601,PhysRevB.104.045122}. Whether universal behavior among kagome metals extends to FeGe remains an open question. 
This motivates us to go beyond modeling via a single-orbital kagome lattice and connect to a realistic band structure~\cite{jiang2023kagome}. Our approach then elucidates the role of breaking spin-SU(2) symmetry by the AFM parent phase on secondary phases in the electronic phase diagram. 

In our instability analysis of FeGe electron bands, we uncover universal aspects connected to the kagome lattice, in addition to qualitatively different effects due to the AFM background. We argue that bare band susceptibilities can be approximated by a 2D projection with VHPs at the $M$~points 
of almost pure sublattice type. On top of  a resulting sublattice interference mechanism, we also find that an on-site Hubbard interaction is also ineffective in coupling opposite spins at (spin-degenerate) VHPs because of the FM layer polarization. As a consequence, transversal spin orders are suppressed in comparison to the SU(2) symmetric phase diagram. 
Furthermore, we find a strong tendency towards spin-triplet $f$-wave pairing. Our analysis places FeGe at the boundary of a $p$-wave CDW, $f$-wave pairing, and inter-sublattice spin density wave instability. In the phase diagram, we also identify a canted antiferromagnetic order which is consistent with experimental observations \cite{Chen2024canting} and a ferrimagnet.

\begin{figure}[t]
    \centering
    \includegraphics[width=1.\columnwidth]{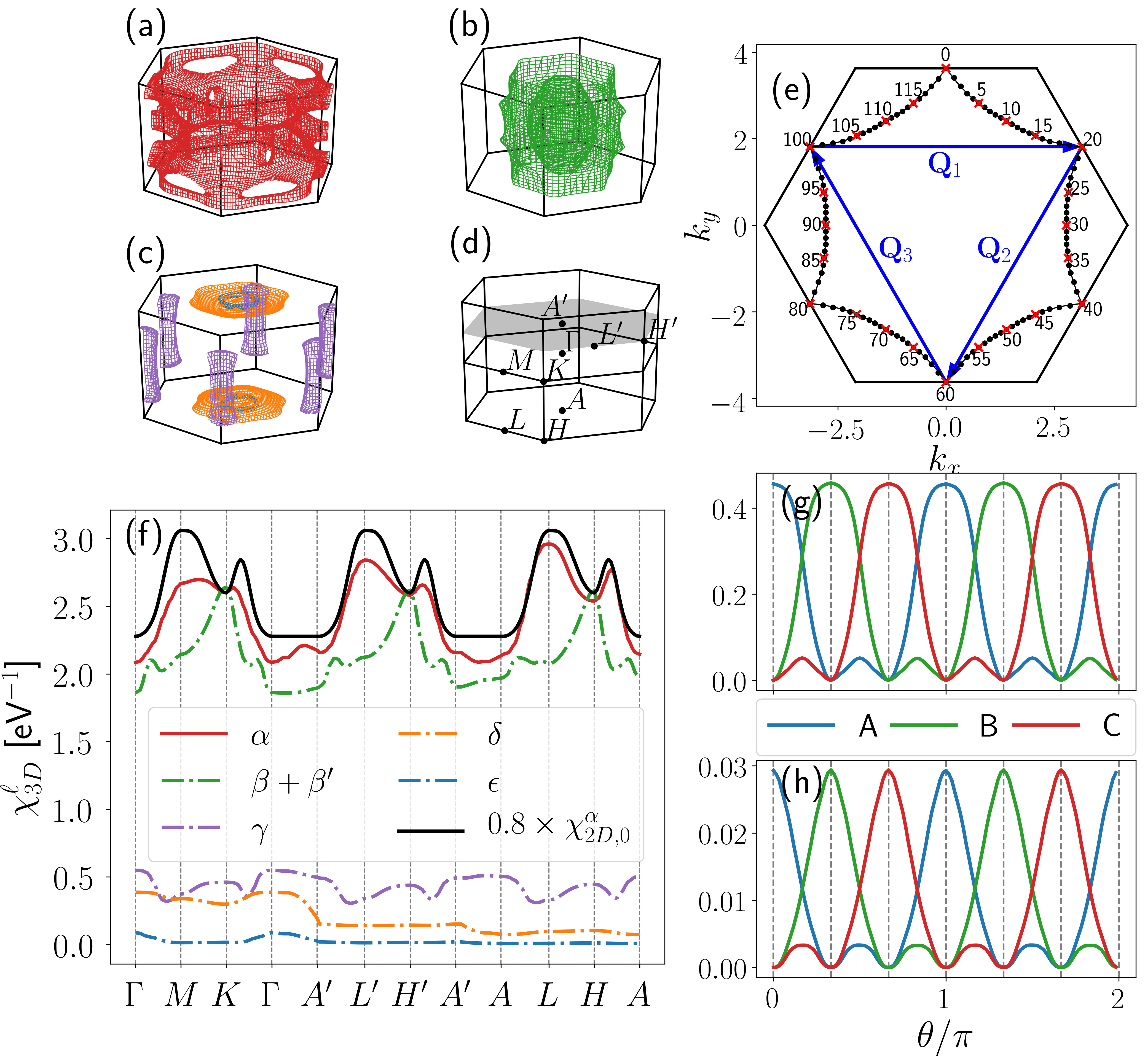}
    \caption{
    Fermi surfaces of the (a)~$\alpha$-band (b)~$\beta$ and $\beta'$ bands, (c)~$\gamma$ (purple), $\delta$ (orange), and $\epsilon$ (blue) bands. (d)~Points in the three-dimensional Brillouin zone. The shaded plane is the $k_z=\pi/2$ plane.
    (e)~Fermi surface and patching scheme used for the FRG calculations for the cut of the $\alpha$-band at $k_z=k_{z,0}\simeq 1.425$ 
    with quasi-2D van Hove singularity. 
    Blue arrows indicate approximate nesting vectors. (f)~3D band susceptibility $\chi_{3D}^\ell(\bm q)$ for the bands crossing the Fermi level and 2D susceptibility $\chi^\alpha_{2D,2}$ at $k_{z,0}\simeq1.425$ (black solid line) at temperature $T=30$~meV.  (g)~and (h): Spin-up Fermi surface orbital weights of the Fe $d_{xz}$ orbitals on the three kagome sublattices on layer 1~(g) and layer 2~(h) for the cut at $k_{z,0}=1.425$. The angle $\theta$ swipes around the Fermi surface. Van Hove points (gray vertical lines) are mainly made of a single sublattice and layer contribution.}
    \label{fig:suscep}
\end{figure}

\paragraph{Band structure and susceptibilities.}~
We perform an \textit{ab-initio} calculation of the band structure of FeGe in the AFM phase as detailed in Ref.~\cite{jiang2023kagome}.  
In the AFM state bands are doubly degenerate due to a generalization of Kramers' theorem based on the combined operation of time-reversal and one-site translation along the $c$-axis $T_z \otimes \mathcal{T}$. 
Consequently, spin and layer are locked. 
There are five bands $E^\ell_\bk$ crossing the Fermi level $\ell\in \{\alpha,\ldots,\epsilon\}$, out of which some have 2D and some have 3D character, as shown in Fig.~\ref{fig:suscep}(a-c) and in~\cite{SM} Sec.~\ref{SMsec: abinitio FS}. We find that the dominant contribution at the Fermi level comes from $d_{xz}, d_{yz}$, and $d_{x^2-y^2}$ Fe orbitals, which strongly contribute to the orbital composition of approximately 2D $\alpha,\beta$, and $\gamma$ Fermi surfaces. 

In order to identify the Fermi surface sheets where intra-band electronic correlation effects are the most important, we compute the band susceptibility $ \chi_{3D}^\ell( \bq) = T\sum_\omega \int_{\mathrm{BZ}}\!d^3\bk\, G^\ell(\omega,\bk+\bq)G^\ell(\omega,\bk)$ for each band crossing the Fermi level, where $G^\ell(k)=(i\omega - E^\ell_\bk)^{-1}$ \cite{footnote_inter}. 
In Fig.~\ref{fig:suscep}(f) we show that the Fermi surface $\alpha$-sheet yields the largest susceptibility 
with a maximum at the $L$-point and 
at related points $M$, $L'$ with the same planar coordinates. Furthermore, we observe a quasi-2D form of the largest band susceptibilities $\chi_{3D}^\ell( \bq)$ in that their qualitative behavior is independent of $q_z$. We observe the analogous 2D behavior in the largest orbital susceptibilities (see~\cite{SM} Secs.~\ref{SMsec: abinitio FS}, \ref{SMsec: 2D single band approx}). 
To find an appropriate 2D approximation, we determine a representative Fermi surface from a cut at fixed $k_{z}=k_{z,0}$ (Fig.~\ref{fig:suscep}(e)) whose in-plane susceptibility $ \chi_{2D}^\alpha(\bq) = T\sum_\omega \int_{\mathrm{2D-BZ}}\!d^2\bk\, G^\alpha(\omega,\bk+\bq,k_{z,0})G^\alpha(\omega,\bk,k_{z,0})$ reproduces the largest band susceptibility $\chi^{\alpha}_{3D}$, Fig.~\ref{fig:suscep}(f). The peaks at M arise due to scattering processes that connect Van Hove points. 

\paragraph{Ab-initio-derived 2D model.}~Based on the analysis of band susceptibilities, we consider an effective model that contains
the $\alpha$ band and its projected interactions in the following FRG calculation. 
We start from the extended Hubbard-Kanamori Hamiltonian that contains spin-dependent intra- and inter-orbital density-density $U_{\sigma\sigma'}^{(\prime)}$ and exchange $J_{\sigma\sigma'}$ couplings for the five Fe $d$-orbitals on a given site, as well as an orbital-independent nearest-neighbor interaction. To project the interaction onto the $\alpha$-band, we approximate the electron operator for sublattice~$b$, orbital~$m$, and spin-layer-locked~$\sigma$ as $c_{\bk,b,m,\sigma} \approx \phi_{\bk,b,m,\sigma} a_{\bk,\sigma}$, where $a_{\bk,\sigma}$ annihilates an electron in the $\alpha$-band with momentum $\bk$ and $ \phi_{\bk,b,m,\sigma}$ is the eigenvector of the \textit{ab-initio} Hamiltonian. We then obtain the projected interaction  
\begin{equation}
\begin{split}
\mathcal{H}_\mathrm{int}&= \int_{\substack{\bk_1,\bk_2\\\bk_3,\bk_4}} \,\sum_{\substack{\sigma_1,\sigma_2\\ \sigma_3,\sigma_4}} V_{\sigma_1\sigma_2\sigma_3\sigma_4}(\bk_1,\bk_2,\bk_3)\\
& \times\delta_{\bk_1+\bk_2-\bk_3-\bk_4,0}\, a^\dagger_{\bk_1,\sigma_1}a^\dagger_{\bk_2,\sigma_2}a_{\bk_4,\sigma_4}a_{\bk_3,\sigma_3}\,,
\end{split}  
\end{equation}
with $\int_\bk=\int_\mathrm{2D-BZ}{d^2k/(2\pi)^2}$. Exploiting the discrete symmetry of the AFM, as well as its residual U(1) symmetry, we can reduce $V_{\sigma_1\sigma_2\sigma_3\sigma_4}(\bk_1,\bk_2,\bk_3)$ to two independent components: $V_{\uparrow\uparrow},V_{\uparrow\downarrow}$ (\cite{SM} Sec.~\ref{SMsec: FS proj}). 

Analyzing the layer, and sublattice content of the Fermi surface cut at $k_z=k_{z,0}$, we find that 
each 2D-like van Hove singularity of the Fermi surface cut at $k_{z,0}$ has weight on only one of the three kagome sublattices, Fig.~\ref{fig:suscep}(g) and \ref{fig:suscep}(h), as it happens in simpler tight binding models~\cite{PhysRevB.86.121105}. In our case, 
the Fermi surface contribution for spin up (down) mostly comes from the first (second) layer (cf. scales of Figs.~\ref{fig:suscep}(g) and \ref{fig:suscep}(h)).

\paragraph{Instability analysis.}~We employ FRG to calculate the dressed interaction of the many-body system 
which takes into account all one-loop particle-particle and particle-hole fluctuations induced by the $\alpha$-band electrons (see~\cite{SM} Sec.~\ref{SMsec: patch fRG} for details). 
For consistency reasons with our approximation of the ab-initio band structure, we resolve the momentum dependence via a patching scheme of the Fermi surface with 120 patch points (Fig.~\ref{fig:suscep}(e))~\cite{RevModPhys.84.299,doi:10.1080/00018732.2013.862020}. 
An instability occurs when the renormalized interaction develops a characteristic singularity
at a mean-field-like critical temperature $T_c$ (e.g., Fig.~\ref{fig:fig3}(g)) and Figs.~\ref{figSM: 8}-\ref{figSM: 9bis}~\cite{SM}. 
From the structure of the singularity on the Fermi surface, we infer 
the order parameter corresponding to the instability 
and the form factors that specify the angular harmonic of the particle-hole or Cooper pair. 
To diagnose if the leading instability occurs in charge~(c), transverse~($x$,$y$), and longitudinal~($z$) spin channel, 
we define  
$V_c(\theta_1,\theta_2,\theta_3) = V_{\up\up}(\theta_1,\theta_2,\theta_3) + V_{\up\down}(\theta_1,\theta_2,\theta_3)$, 
$V_x(\theta_1,\theta_2,\theta_3) = V_y(\theta_1,\theta_2,\theta_3) = - V_{\up\down}(\theta_2,\theta_1,\theta_3)$, 
and $V_z(\theta_1,\theta_2,\theta_3) = V_{\up\up}(\theta_1,\theta_2,\theta_3) - V_{\up\down}(\theta_1,\theta_2,\theta_3)$. 
As in the experiments, the $z$-axis is chosen along the alignment of the inter-plane antiferromagnetic order parameter.
%
\begin{figure}[t]
    \centering
    \includegraphics[width=1.\columnwidth]{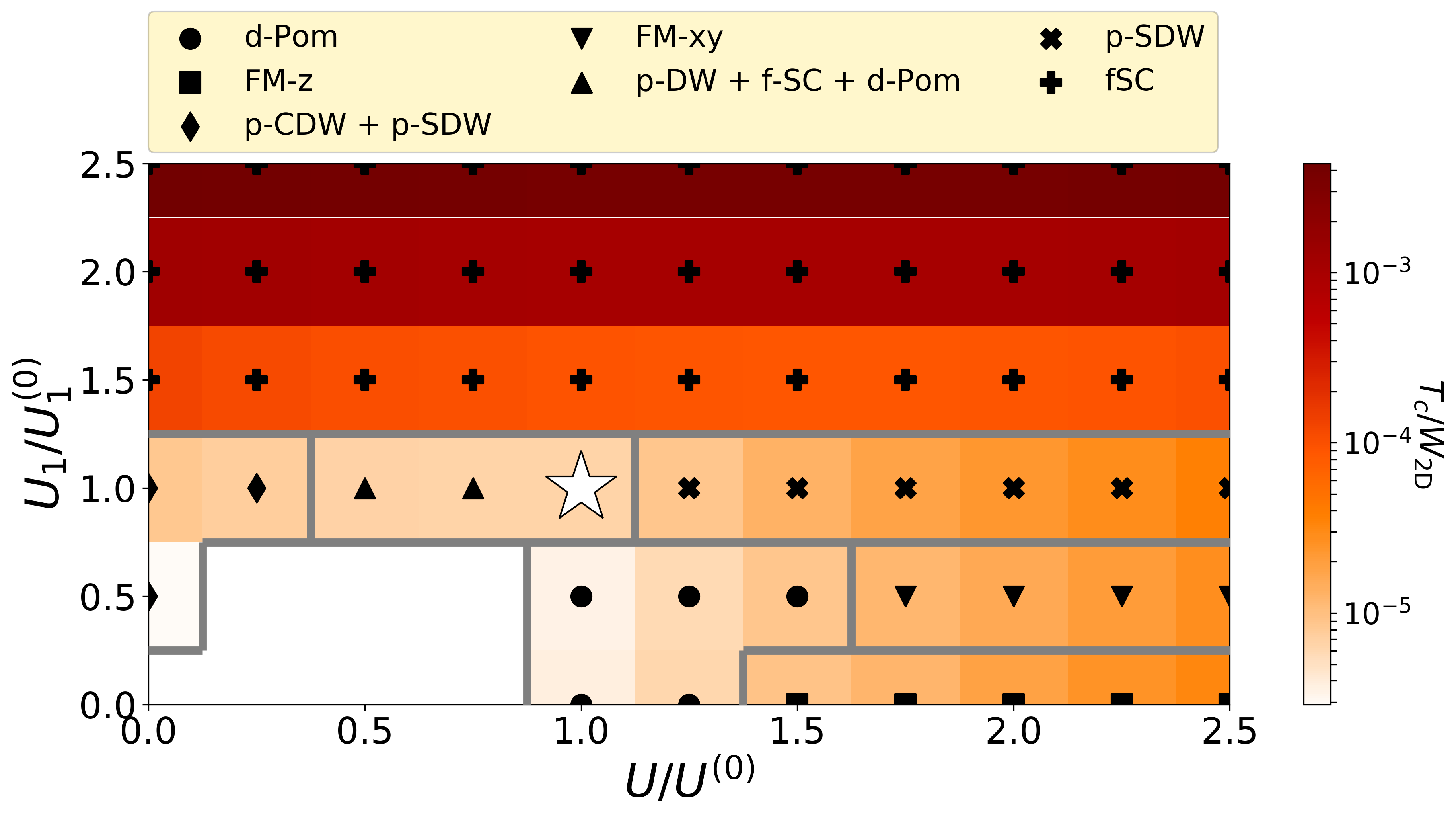}
    \caption{Phase diagram and FRG critical temperatures $T_c$ of \textit{ab-initio}-derived, quasi-2D approximation for FeGe as a function of the onsite Hubbard $U$ and nearest-neighbor interaction $U_1$ in units of their cRPA values $U^{(0)}=5.72$ eV and $U_1^{(0)}=2.30$ eV for fixed cRPA $U'_{\sigma,\sigma'}$ and $J_\sigma$. The band width is $W_{2D}\sim 450$ meV$\sim 5000$ K. 
    Abbreviations mean the following. FM-z: ferrimagnetism along the $z$-axis, p-CDW+p-SDW: $p$-wave charge and spin density wave order, d-Pom: $d$-wave spin Pomeranchuk, FM-xy: canted antiferromagnetism, and f-SC: $f$-wave superconductivity. 
    The white star marks the point in the phase diagram corresponding to cRPA values of the interactions.}
    \label{fig:pd}
\end{figure}
\begin{figure*}[t]
    \centering
    \includegraphics[width=0.8\textwidth]{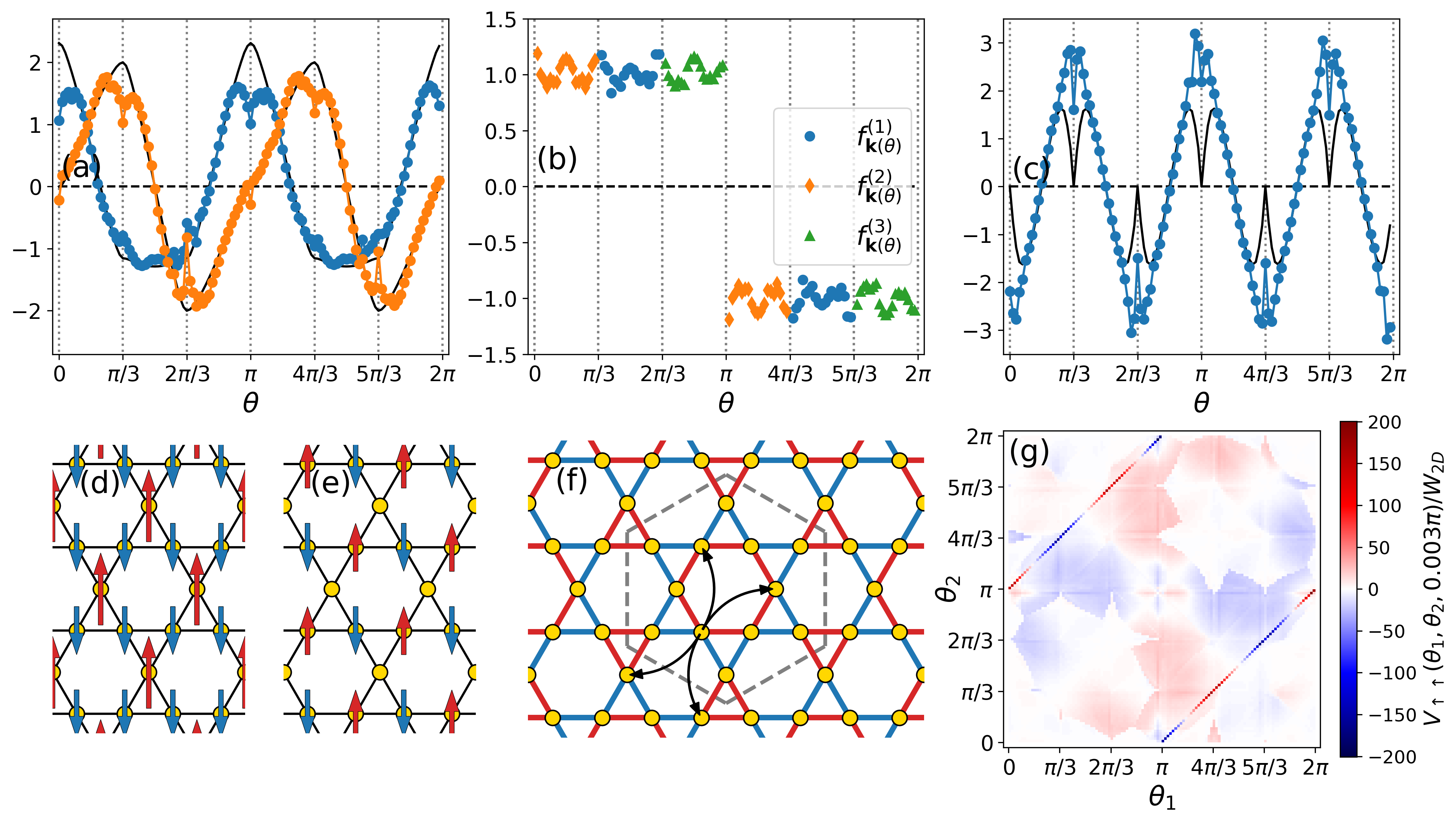}
    \caption{(a) Leading (degenerate) eigenfunctions of $V_z(\theta_1,\theta_2,\theta_1)$ in the spin-Pomeranchuk phase. Their angular dependence is well approximated by the form factors $d^{x^2-y^2}_\bk$ and $d^{xy}_\bk$ 
    (solid black lines) with 
    form factors defined in the main text  (b) Angular dependence of the leading vertex eigenfunction in the $p$-CDW/SDW phase for $\bQ=\bQ_1$ (blue dots), $\bQ=\bQ_2$ (orange diamonds), and $\bQ=\bQ_3$ (green triangles). (c) Angular dependence of the leading eigenfunction of $V_{\up\up}(\theta_1,(\theta_1+\pi)\mathrm{mod} (2\pi),\theta_2)$ in the $f$-wave superconducting phase. The function is well approximated by 
    $(\sin(\sqrt{3}k_y)-2\sin(3k_x/2))\cos(\sqrt{3}k_y/2)$ (solid black line). 
    (d) $d_{x^2-y^2}$ and (e) $d_{xy}$ representative spin patterns of the spin-Pomeranchuk phase. Spins pointing upwards in (d) are twice as long as those pointing downwards. Note that these patterns only depict the deviation from the underlying interlayer antiferromagnetic background. (f) Charge/spin bond density patterns in the $p$-CDW/SDW phase. A red (blue) bond represents an enhanced (suppressed) hopping along that bond. 
    The enlarged unit cell  
    contains 12 kagome sites (gray dashed line). Black arrows connect next-nearest-neighbor sites coupled by $f$-wave pairing. (g) Typical interaction vertex $V_{\uparrow\uparrow}(\theta_1,\theta_2,\theta_3\simeq 0)$ between equal spin electrons 
    for the $f$-wave pairing instability.}
    \label{fig:fig3}
\end{figure*}

\paragraph{Phase diagram.}~Based on the instability analysis, we determine the phase diagram as function of on-site $U$ and nearest-neighbor $U_1$ interaction shown in Fig.~\ref{fig:pd}. For small values of $U$ and $U_1$, we find no instability down to the lowest numerically resolved temperature, $T\sim 10$~mK. 
For large enough values of $U\gtrsim U^{(0)}=5.72$eV or $U_1\gtrsim U_1^{(0)}=2.30$eV, we obtain different types of instabilities discussed below. They evolve from 
$q=0$ spin/charge instabilities (ferrimagentism, canted antiferromagnetism, or d-wave spin Pomeranchuk) over $q=M$ density-waves to $f$-wave pairing instabilities 
upon increasing $U_1$. Throughout the phase diagram, we obtain small critical temperatures $T_c$ as compared to the two-dimensional bandwidth, which signals strongly competing orders. 
We find that $T_c$ depends only mildly on the interaction $U$ and increases with $U_1$. 

The $q=0$ instabilities spontaneously break the combined $T_z \otimes \mathcal{T}$ of the AFM. Among them, we find longitudinal ($i=z$) and transversal ($i=x,y$) ferrimagnetic (FM) orders  for $U\gtrsim 1.5 U^{(0)}$, $U_1\lesssim U_1^{(0)}$ with order parameter 
\begin{equation}\label{seq: p-DW OPs}
    S^i = \int_{\bk\in\mathrm{FS}}\langle a^\dagger_\bk\, \sigma^i \,a_{\bk}\rangle \,.
\end{equation}  
The transversal FM also breaks the residual U(1), the longitudinal does not. The longitudinal FM describes a magnetic phase in which the spins in neighboring layers point in opposite directions but with different magnitudes. Moreover, the two layers have different average charge densities. In the transversal FM, the spins get a canting pointing in the same direction in both layers. 
For $U\lesssim 1.5U^{(0)}$, we find a $d$-wave spin Pomeranchuk order 
\begin{equation}\label{seq: p-DW OPs}
    S^z_n = \int_{\bk\in\mathrm{FS}}
    d_\bk^n\langle a^\dagger_\bk\, \sigma^z \,a_{\bk}\rangle \,,
\end{equation}  
in close competition with the corresponding charge instability. Due to the lattice $D_{6h}$ symmetry, there are two degenerate  form factors 
%
$d^{x^2-y^2}_\bk = (2/\sqrt{3})[\cos(k_x) -\cos(k_x/2)\cos(\sqrt{3}k_y/2)]$ and $d^{xy}_\bk=2\sin(k_x/2)\sin(\sqrt{3}k_y2)$, 
Fig.~\ref{fig:fig3}(a). The order additionally breaks 
$D_{6h}$ 
spontaneously when a specific linear combination of them, 
develops below the critical temperature, i.e., it is a spin-polarized nematic \cite{PhysRevB.75.115103,gali2024,PhysRevB.107.125142,PhysRevB.102.125141,PhysRevB.102.125120}. The two independent $d_{xy}$- and $d_{x^2-y^2}$-representatives have the same critical temperature but can have different energies in the symmetry-broken state. After a Fourier transform, the spin-Pomeranchuk phase manifests as an inter-sublattice spin-density wave in real space.  
The two corresponding patterns in Fig.~\ref{fig:fig3}(d-e) 
represent an additional modulation of the charge density in each layer, which is proportional to the local spin amplitude. It has to be added (subtracted) from the spin magnitude in the top (bottom) layer. 

For $U_1$$\approx U_1^{(0)}$ 
and a range of $U$, we find that the system is unstable towards a $p$-wave density wave with $q=\bQ_1=(2\pi,0)$, $q=\bQ_2=(-\pi,-\sqrt{3}\pi)$, and $q=\bQ_3=(-\pi,\sqrt{3}\pi)$
(Fig.~\ref{fig:suscep}(e)), where $\bQ=M+G$ with G a reciprocal lattice vector. The order parameter reads:
\begin{equation}\label{seq: p-DW OPs}
    \Phi_n = \int_{\bk\in\mathrm{FS}} f^{(n)}_\bk\langle a^\dagger_\bk\, \eta \,a_{\bk+\bQ_n}\rangle \,,
\end{equation} 
where $\eta=\mathbb{1}$ or $\sigma^z$, and $f^{(n)}_\bk=\sin(\bk\cdot\bQ_n/(4\pi))$~\cite{footnoteQn}. Such a phase corresponds to a charge or spin bond density wave and breaks translation symmetry. It also breaks  $T_z \otimes \mathcal{T}$ symmetry if $\eta=\sigma_z$. It is analogous to instabilities  found in paramagnetic Hubbard models on a kagome lattice~\cite{PhysRevLett.110.126405,PhysRevB.87.115135,profe2024kagome}. In Fig.~\ref{fig:fig3}(b) we show the computed momentum dependence of the functions $f^{(n)}_\bk$ for those combinations $\bk$ and $\bk+\bQ_n$ that are close to the Fermi surface. 
In the ordered state, one or more of the order parameters in Eq.~\eqref{seq: p-DW OPs} can condense. 
To determine the ground state configuration, we consider the Landau free energy 
\begin{equation}
    \begin{split}
        V[\Phi_1,\Phi_2,\Phi_3] = &r\sum_{n=1}^3 \Phi_n^2+u\left(\sum_{n=1}^3\Phi_n^2\right)^2\\
        &+\lambda\left(\Phi_1^2\Phi_2^2+\Phi_2^2\Phi_3^2+\Phi_3^2\Phi_1^2\right)\,.
    \end{split}
\end{equation}
In \cite{SM} Sec.~\ref{SMsec: phase diagram}, we calculate 
the parameter $\lambda$ from the band dispersion $E^\alpha(\bk)$, obtaining $\lambda<0$ , which means that a state with all $\Phi_n$ non-zero and equal in magnitude minimises the energy. We show the corresponding real-space charge or spin bond density patterns in Fig.~\ref{fig:fig3}(f).

Finally, for $U_1\geq 1.5 U_1^{(0)}$ independent of the value of $U$, we find a superconducting pairing instability in $V_{\uparrow\uparrow}$ indicating spin-triplet pairing of equal spins. We infer an order parameter
\begin{equation}\label{seq: p-DW OPs II}
    \Delta = \int_{\bk\in\mathrm{FS}} f^f_\bk \langle a_\bk\, (\cos\theta \sigma_x+e^{i\varphi}\sin\theta\sigma_y)i\sigma_y \, a_{-\bk}\rangle \,,
\end{equation} 
parameterized by $\theta$ and $\varphi$ with $f$-wave symmetry $f^f_\bk \propto \sin(\sqrt{3}k_y) -2\cos(3k_x/2)\sin(\sqrt{3}k_y/2)$ in the $B_{2u}$ irreducible representation of $D_{6h}$. 
To understand the pairing mechanism better, we analyze the bare pairing vertex and find that there is a small attractive $f$-wave component after projection into the $\alpha$ band. It induces a pairing instability through enhancement by fluctuations in the particle-hole channels. 
This pairing state couples electrons of equal spin on next-nearest-neighbor sites (see arrows in Fig.~\ref{fig:fig3}(f)). 
In the AFM phase of FeGe equal spin pairing from electronic interactions is expected to be favored due to an almost complete layer polarization, which strongly suppresses interactions between unequal spin electrons. 
More generally, almost all the phases we found in our computations are described by order parameters that couple equal-spin electrons, the only exception being the $xy$ canted antiferromagnetic phase. 

\paragraph{Conclusion.}~ 
We studied ordering tendencies of the magnetic kagome metal FeGe driven by electronic correlations. 
To that end, we employed FRG as an unbiased many-body technique starting from a 2D approximation for realistic DFT band structures.
Our analysis reveals a strong competition of orders due to the interplay of spin, charge, and pairing correlations. 
The $p$-wave CDW with wave vector $M$ and the canted AFM phase we found are consistent with experimental observations of the CDW \cite{Teng2022,Teng2023,PhysRevResearch.6.013276,PhysRevLett.132.266505,yi2024polarizedcharge,shi2024,oh2024tunability,PhysRevLett.133.046502,han2024,subires2024} and canted phase \cite{footnote,Chen2024canting,doi:10.1143/JPSJ.18.589,Beckman_1972,Bernhard_1984,Bernhard_1988}.

From a qualitative point of view, we find clear signatures of electronic behavior characteristic for the kagome lattice. 
We reveal sublattice interference in an effective, \emph{ab-initio} inspired model for FeGe, which we expect to be the reason behind similarities of our phase diagram with that of the pure kagome lattice Hubbard model~\cite{PhysRevLett.110.126405,PhysRevB.87.115135,profe2024kagome,schwemmer2024,PhysRevB.86.121105}. Qualitative similarities include q=0 magnetic instabilities at small, charge/spin-bond order for intermediate, and pairing instabilities for large nearest-neighbor interaction. This constitutes a non-trivial outcome given the different starting points and exposes universal aspects between magnetic and non-magnetic kagome metals. Taking a recent comparison~\cite{profe2024kagome} between different FRG schemes for the momentum resolution into account, we expect this conclusion to remain qualitatively correct, even if exact phase boundaries may shift. 

However, there are also qualitative differences which we mainly attribute to the broken SU(2) of the AFM parent state in FeGe. We find that this promotes longitudinal over transversal spin orders because the on-site interaction cannot couple electrons with opposite spins even though they have degenerate energies (the "Kramers doublet" is nonlocal due to $T_z \otimes T$ symmetry). Furthermore, we find a strong tendency towards $f$-wave triplet superconductivity. While $f$-wave pairing was discussed for the paramagnetic kagome lattice in the context of phonons, disorder, or doping \cite{PhysRevB.108.144508,PhysRevLett.110.126405,PhysRevB.109.014517}, here it becomes a very robust instability for a slight increase of the nearest-neighbor interaction. This opens the intriguing possibility of inducing superconductivity also in magnetic kagome metals if the ratio of nearest-neighbor over on-site interaction can be tuned, e.g., via pressure or doping \cite{korshunov2024pressure}. 

Our results also indicate that a purely electronic mechanism cannot explain the occurrence of the charge density wave at relatively high temperatures of 100~K. 
Thus it would be very interesting to include the cooperation with phonons \cite{PhysRevMaterials.8.L080601,Wu_2023,Ma_2024}. At the same time, the transition temperature for long-range CDW order is significantly reduced in annealed samples \cite{PhysRevLett.132.256501,oh2024tunability,Tan2024}. Generally, it will be crucial to include other bands when studying orders that compete with a CDW and further work out universal aspects of the Kagome metals. This can be done, e.g., by analyzing AFM kagome Hubbard models or by determining the CDW landscape of phenomenological Landau models for DFT results.

\paragraph{Acknowledgments.}~We thank Morten H. Christensen, Maria N. Gastiasoro, Walter Metzner, Subir Sachdev, Tomohiro Soejima, and Roser Valent\'i for discussions, and Walter Metzner for a critical reading of the manuscript. M.M.S. acknowledges funding from the Deutsche Forschungsgemeinschaft (DFG, German Research Foundation) within Project-ID 277146847, SFB 1238 (project C02), and the DFG Heisenberg programme (Project-ID 452976698). P.M.B. acknowledges support by the German National Academy of Sciences Leopoldina through Grant No. LPDS 2023-06 and funding from U.S. National Science Foundation grant No. DMR2245246. 
Y.J. and H.H were supported by the European Research Council (ERC) under the European Union’s Horizon 2020 research and innovation program (Grant Agreement No. 101020833), as well as by the IKUR Strategy under the collaboration agreement between Ikerbasque Foundation and DIPC on behalf of the Department of Education of the Basque Government. 
DC acknowledges support from the DOE Grant No. DE-SC0016239. 
BAB was supported by the Gordon and Betty Moore Foundation through Grant No. GBMF8685 towards the Princeton theory program, the Gordon and Betty Moore Foundation’s EPiQS Initiative (Grant No. GBMF11070), the Office of Naval Research (ONR Grant No. N00014-20-1-2303), the Global Collaborative Network Grant at Princeton University, the Simons Investigator Grant No. 404513, the BSF Israel US foundation No. 2018226, the NSF-MERSEC (Grant No. MERSEC DMR 2011750), the Simons Collaboration on New Frontiers in Superconductivity, and the Schmidt Foundation at the Princeton University. 
LC was supported by a grant from the Simons Foundation SFI-MPS-NFS-00006741-11. 

\bibliography{main.bib}

\begin{thebibliography}{10}
\bibitem{footnoteQn}
\bibinfo{title}{{In our gauge choice for the orbital-to-band transformation~\cite{SM} the form factors are not periodic and it is important to distinguish Q and M in the transformation to real space.}}
\bibitem{footnote}
\bibinfo{title}{{In our quasi-2D model, an incommensurate, out-of-plane magnetic wave vector $(\pi/2+\delta)\hat e_z$ that connects minority and majority spin would manifest as a $q=0$ FM-xy order.}}
\end{thebibliography}

\newpage


\section*{Supplementary material}

# Supplemental material on: Competing phases in kagome magnet FeGe from functional renormalization

Pietro M. Bonetti,[1,2] Yi Jiang,[3] Haoyu Hu,[4] Dumitru Călugăru,[4]
Michael M. Scherer,[5] B. Andrei Bernevig,[4,3,6] and Laura Classen[1,7]

[1]*Max Planck Institute for Solid State Research, Heisenbergstrasse 1, D-70569 Stuttgart, Germany*
[2]*Department of Physics, Harvard University, Cambridge MA 02138, USA*
[3]*Donostia International Physics Center (DIPC),
Paseo Manuel de Lardizabal. 20018, San Sebastian, Spain*
[4]*Department of Physics, Princeton University, Princeton, NJ 08544, USA*
[5]*Theoretische Physik III, Ruhr-Universität Bochum, D-44801 Bochum, Germany*
[6]*IKERBASQUE, Basque Foundation for Science, 48013 Bilbao, Spain*
[7]*School of Natural Sciences, Technische Universität München, 85748 Garching, Germany*

(Dated: November 17, 2024)

## SI. FERMI SURFACES IN THE AFM PHASE

We performed an *ab-initio* calculation of the band structure of FeGe in the antiferromagnetic (AFM) phase as detailed in Ref. [1]. We use the Vienna *ab-initio* Simulation Package (VASP)[2–6] to obtain the band structure, and *Wannier90*[7–10] to obtain *ab-initio* Wannier tight-binding model based on maximally localized Wannier functions (MLWFs). In the AFM phase, each layer becomes spin polarized with opposite polarization in neighboring layers.

In Fig. S1, we show the 3D Fermi surface of FeGe in the AFM phase. In Fig. S2, we show different Fermi surface cuts for different values of $k_z$, measured in units of the inverse *c*-axis spacing, ranging from 0 to $\pi$. We observe several Fermi surface sheets: two small $\gamma$-pockets (violet dash dotted lines) centered around the $K$-point that extend up to $|k_z| \sim 2.094$, an approximately cylindrical (green dash dotted lines) $\beta'$-sheet centred around the $\Gamma$-point and extending up to $|k_z| \sim 2.513$, an open $\beta$-sheet (green dash dotted lines) in the form of a hexagonal prism, crossing the $k_z = \pi$ plane, and a large Fermi surface (red solid line) that changes character from hole-like to electron-like several times as a function of $k_z$. An additional 'disk-shaped' (orange dash dotted lines) $\delta$ Fermi surface, as well as an $\epsilon$ 'ring-shaped' (blue dash dotted lines) one appear in the vicinity of $k_z = \pi$. In Fig. S3, we show the same Fermi surfaces as in Fig. S2 but along cuts performed at $k_y = 0$ (panel a), $k_x = 0$ (panel b), and $k_y = (2\pi)/\sqrt{3}$ (panel c), the latter corresponding to a side face of the three-dimensional BZ. The color code is the same as in Fig. S2. In panels (b) and (c) the nature of the Fermi surface marked by a red solid line is evident: at the $k_z$ values marked by horizontal gray dash dotted lines we have a change in topology in the two dimensional cut of the Fermi surface.

We would like to analyze if the CDW in FeGe can arise as a Fermi-surface instability because of experimental signatures that the CDW wave vector is $M$ or $L$ and connects Van Hove points at the boundary of the Brillouin zone. A random phase approximation (RPA) in the charge-density-wave channel can be performed either in band or orbital basis, and will generally couple different bands or orbitals, i.e., one cannot directly read of the instability at a single band/orbital. To identify processes with large contributions in the different bases, we compute band and orbital susceptibilities. We define the band susceptibility for bands crossing the Fermi level as

$$\chi_{0,band}^{b,b'}(\boldsymbol{q}) = \int_{\mathrm{BZ}} \frac{d^3\mathbf{k}}{(2\pi)^3} \frac{f(E_{\mathbf{k+q}}^b) - f(E_{\mathbf{k}}^{b'})}{E_{\mathbf{k}}^b - E_{\mathbf{k+q}}^{b'}} \,, \tag{S1}$$

with $f(x) = (1 + e^{x/T})^{-1}$ the Fermi function and $E_{\mathbf{k}}^b$ the dispersion of the $b$-th band. In Fig. S4 we show that the Fermi surface $\alpha$-sheet (indicated by a solid red line in Fig. S2 and S3) has the largest intra-band susceptibility ($b = b'$) compared to all other bands crossing the Fermi level. Furthermore, we observe that the $\alpha$-band susceptibility exhibits a weak dependence on $q_z$, with peaks occurring at the $L$-point $\mathbf{k} = (0, (2\pi)/\sqrt{3}, \pi)$ and related points with the same in-plane coordinates. The largest inter-band susceptibility comes from processes connecting $\alpha, \beta$ bands. The $k_z$ dependence is also very weak and peaks occur at points which project onto K points in the plane, Fig. S5.

We also consider the intra-orbital susceptibility defined as [11]

$$\chi_0^{o,s,l,\sigma}(\mathbf{q}) = \int_{\mathrm{BZ}} \frac{d^3\mathbf{k}}{(2\pi)^3} \sum_{b,b'} |\phi_{\mathbf{k},o,s,l,\sigma}^b|^2 |\phi_{\mathbf{k+q},o,s,l,\sigma}^{b'}|^2 \frac{f(E_{\mathbf{k+q},\sigma}^{b'}) - f(E_{\mathbf{k},\sigma}^b)}{E_{\mathbf{k},\sigma}^b - E_{\mathbf{k+q},\sigma}^{b'}} \,, \tag{S2}$$

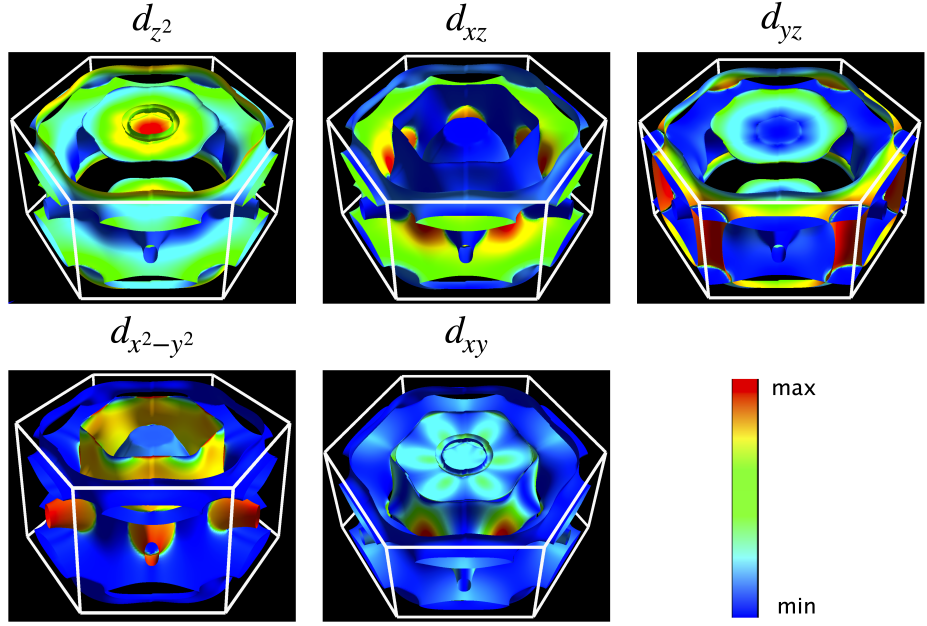

Figure S1. The Fermi surface of FeGe in the AFM phase. In each panel we show the Fe orbital weight on the Fermi surface with a color scale. The respective orbital is indicated on top of each panel. Note that some sheets of the Fermi surfaces with negligible orbital weights are omitted for clarity and improved visualization.

where $\sigma = \uparrow, \downarrow$ labels the spin, $o \in \{d_{z^2}, d_{xz}, d_{yz}, d_{x^2-y^2}, d_{xy}\}$ the Fe $d$-orbitals, $s \in \{A,B,C\}$ the kagome sublattice, $l \in \{1, 2\}$ the layer, and $b = 1, \ldots, 48$ the bands (from 6 Fe atoms with 5 d-orbitals and 6 Ge atoms with 3 p-orbitals in the AFM unit cell). Here, $\phi^b_{\mathbf{k},o,s,l,\sigma}$ is the transformation matrix from the microscopic basis (parametrized in terms of orbital, sublattice and layer sites) to the band basis. Note that, as the spin-orbital coupling (SOC) is negligible in FeGe, we can treat two spin sectors separately. Two spin sectors share the same dispersion with the wavefunctions related by the anti-unitary translation that connects two layers, that is, $\phi^b_{\mathbf{k},o,s,l,\sigma} = -\text{sign}(\sigma) \phi^b_{\mathbf{k},o,s,\bar{l},\bar{\sigma}}$, with $\bar{1} = 2$, $\bar{2} = 1$, $\bar{\uparrow} = \downarrow$, $\bar{\downarrow} = \uparrow$, $\text{sign}(\uparrow) = +1$, and $\text{sign}(\downarrow) = -1$. In Fig. S6, we compute the susceptibility for all the $d$ orbitals of Fe. We observe again that the $k_z$ dependence is weak. The largest contributions come from $d_{xz}, d_{yz}$ and $d_{x^2-y^2}$ orbitals with peaks at $M$ and $L$ points.

Combining the complementary information from orbital and band susceptibilities, we conclude that the leading contritbutions come from $\alpha$ and $\beta$ bands, whose orbital weight mainly comes from $d_{x^2-y^2}$ and $d_{xz}$ orbitals. Furthermore, susceptibilities can be approximated as being quasi-2D with a negligible dependence on $k_z$.

For concreteness, we now consider the RPA analysis in band space for a general CDW vertex with wave vector $Q$ between bands $b, b'$, $H_{CDW} = \int_\mathbf{k} \Phi^{bb'}_\mathbf{k}(\mathbf{q}) \sum_\sigma c^\dagger_{\mathbf{k}+\mathbf{q},b,\sigma} c_{\mathbf{k},b'\sigma}$+h.c. Within RPA, the self-consistent equation for the CDW vertex is of the form

$$\Phi^{b_1 b_2}_\mathbf{k}(\mathbf{q}) = \Phi^{0,b_1 b_2}_\mathbf{k}(\mathbf{q}) + \sum_{b,b'} \int_{\mathbf{k}'} W^{b_1 b_2 bb'}_{\mathbf{k},\mathbf{k}'}(\mathbf{q}) \, T \sum_\omega G_b(\mathbf{k}' - \mathbf{q}, \omega) G_{b'}(\mathbf{k}', \omega) \Phi^{bb'}_{\mathbf{k}'}(\mathbf{q}) \tag{S3}$$

where $\Phi^{0,b_1 b_2}_\mathbf{k}$ is the bare CDW vertex, $\int_q$ is a shorthand for $\int_{BZ} \frac{d^3 q}{(2\pi)^3}$, $G_b(\mathbf{k}, \omega) = (i\omega - E^b_\mathbf{k})^{-1}$ is the Green's function for a band with dispersion $E^b_\mathbf{k}$, $T \sum_\omega$ is the sum over Matsubara frequencies, and $W^{b_1 b_2 bb'}_{\mathbf{k},\mathbf{k}'}(\mathbf{q})$ is the interaction in the CDW channel $W^{b_1 b_2 bb'}_{\mathbf{k},\mathbf{k}'}(\mathbf{q}) = \frac{1}{2} \sum_\sigma [V^{b_2 bb' b_1}_{\sigma\bar\sigma\sigma\sigma}(\mathbf{k}+\mathbf{q},\mathbf{k}'-\mathbf{q},\mathbf{k}',\mathbf{k}) - V^{b_2 bb_1 b'}_{\sigma\bar\sigma\sigma\sigma}(\mathbf{k}+\mathbf{q},\mathbf{k}'-\mathbf{q},\mathbf{k},\mathbf{k}') - V^{b_2 bb_1 b'}_{\sigma\bar\sigma\bar\sigma\sigma}(\mathbf{k}+\mathbf{q},\mathbf{k}'-\mathbf{q},\mathbf{k},\mathbf{k}')]$ with $\bar\sigma = -\sigma$. The general interaction in band space

$$\hat{V} = \sum_{\mathbf{k}_1,\ldots \mathbf{k}_3} \sum_{b_1 \ldots b_4} \sum_{\sigma_1 \ldots \sigma_4} V^{b_1 \ldots b_4}_{\sigma_1 \ldots \sigma_4}(\mathbf{k}_1, \ldots \mathbf{k}_4) \, \delta_{\mathbf{k}_1 + \mathbf{k}_2 - \mathbf{k}_3 - \mathbf{k}_4} \, c^\dagger_{\mathbf{k}_1 b_1 \sigma_1} c^\dagger_{\mathbf{k}_2 b_2 \sigma_2} c_{\mathbf{k}_4 b_4 \sigma_4} c_{\mathbf{k}_3 b_3 \sigma_3}, \tag{S4}$$

contains the orbital information via transformation factors

$$V^{b_1 \ldots b_4}_{\sigma_1 \ldots \sigma_4}(\mathbf{k}_1, \ldots \mathbf{k}_4) = \sum_{\alpha_i} \phi^{b_1 *}_{\mathbf{k}_1,\alpha_1,\sigma_1} \phi^{b_2 *}_{\mathbf{k}_2,\alpha_2,\sigma_2} \phi^{b_3}_{\mathbf{k}_3,\alpha_3,\sigma_3} \phi^{b_4}_{\mathbf{k}_4,\alpha_4,\sigma_4} U^{\alpha_1 \ldots \alpha_4}(\mathbf{k}_1, \ldots \mathbf{k}_4), \tag{S5}$$

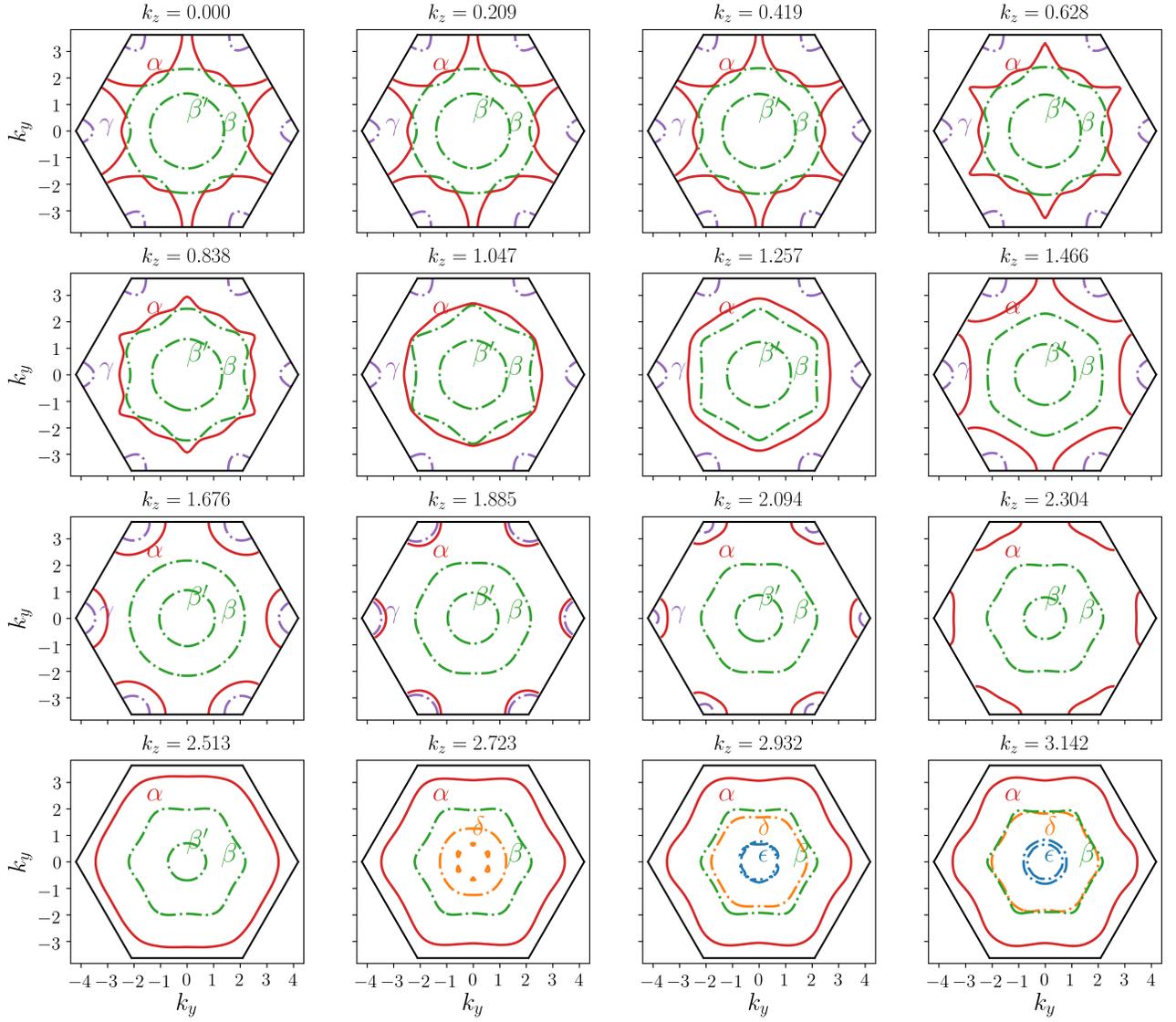

Figure S2. Fermi surface cuts at different values of $k_z$. The value of $k_z$ is indicated on top of each panel.

where $U^{\alpha_1\ldots\alpha_4}(\mathbf{k}_1,\ldots\mathbf{k}_4)$ is the interaction in the microscopic basis and we introduced the muti-index $\alpha = (o,s,l)$ for shortness. The eigenvalues of $W^{b_1 b_2 b b'}_{\mathbf{k},\mathbf{k}'}(\mathbf{q})\, T\sum_\omega G_b(\mathbf{k}'-\mathbf{q},\omega) G_{b'}(\mathbf{k}',\omega)$ viewed as function of both, momentum $\mathbf{k}$ and band $b$ indices, will determine the leading CDW instability. The diagonalisation of Eq. (S3) with respect to momentum can be treated, among others, by multiple insertions of the resolution of the identity $\delta_{\mathbf{k},\mathbf{k}'} = \sum_\ell (f^\ell_\mathbf{k})^* f^\ell_{\mathbf{k}'}$, with $f^\ell_\mathbf{k}$ a suitably chosen set of form factors. We get

$$\hat{\Phi}^{b_1 b_2}_l(\mathbf{q}) = \hat{\Phi}^{0,b_1 b_2}_l(\mathbf{q}) + \sum_{b,b'}\sum_{nm} \hat{W}^{b_1 b_2 b b'}_{ln}(\mathbf{q}) \hat{\chi}^{bb'}_{0,nm}(\mathbf{q}) \Phi^{bb'}_{\mathbf{k}'}(\mathbf{q}) \tag{S6}$$

with $\hat{\Phi}^{bb'}_l(\mathbf{q}) = \int_\mathbf{k} f^\ell_\mathbf{k} \Phi^{bb'}_\mathbf{k}(\mathbf{q})$ and $\hat{W}^{b_1\ldots b_4}_{\ell\ell'}(\mathbf{q}) = \int_{\mathbf{k},\mathbf{k}'} W_{\mathbf{k}\mathbf{k}'}(\mathbf{q}) f^\ell_\mathbf{k} (f^{\ell'}_{\mathbf{k}'})^*$. A similar relation holds for $\hat{\Phi}^{0,bb'}_\ell(\mathbf{q})$. We also defined

$$\chi^{bb'}_{0,\ell\ell'}(\mathbf{q}) = \int_\mathbf{k} (f^\ell_\mathbf{k})^* f^{\ell'}_\mathbf{k}\, \frac{f(E^{b'}_{\mathbf{k}+\mathbf{q}}) - f(E^b_\mathbf{k})}{E^b_\mathbf{k} - E^{b'}_{\mathbf{k}+\mathbf{q}}}\,. \tag{S7}$$

Typically, it suffices to only include a small number of form factors and the expansion can be truncated. If the form factors are chosen according to the lattice symmetries, the form-factor projected susceptibility is diagonal in these

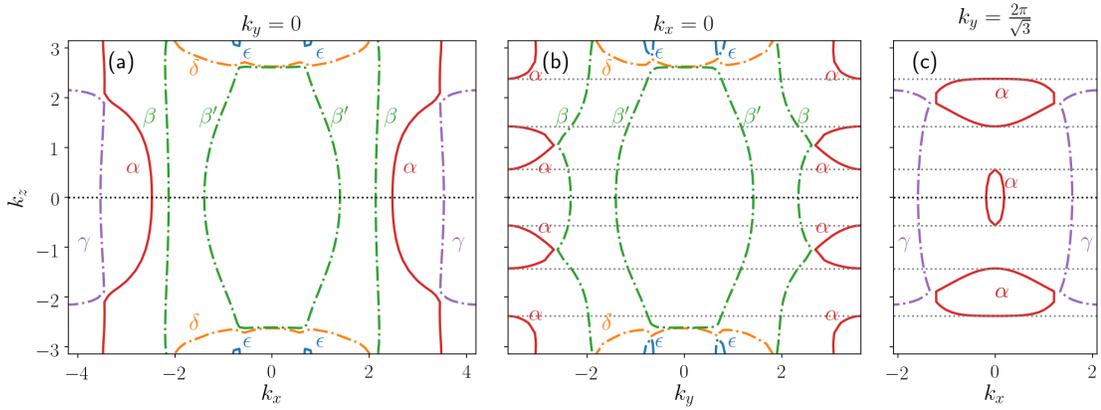

Figure S3. Fermi surface cuts on the $k_y = 0$ (panel a), $k_x = 0$ (panel b), and $k_y = (2\pi)/\sqrt{3}$ (panel c) planes. The color coding is the same as in Fig. S2.

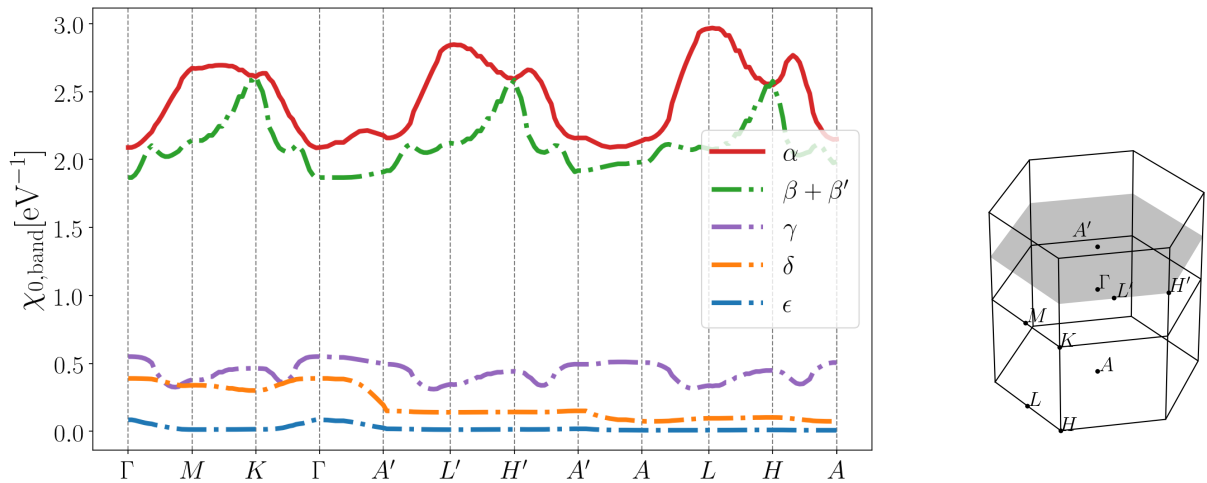

Figure S4. Band susceptibility for the bands crossing the Fermi level. The color coding is the same as in Figs. S2 and S3. The calculations have been performed at the temperature $T = 30$ meV. The units on the $y$-axis are eV$^{-1}$. The points used for the path in the left panel are displayed in the right panel, with the shaded plane indicating the $k_z = \frac{\pi}{2}$ plane.

lowest form factors if $\mathbf{q}$ is a high-symmetry point. As a result we have decoupled the orbital information (which is contained in $\hat{W}$ via the transformation factors $\phi^b_{\mathbf{k},\alpha,\sigma}$) and Fermi-surface effects (contained in $\chi_0$). In Fig. S7 we show several diagonal form factor projected band susceptibilities $\chi^{bb}_{0,\ell\ell}(\mathbf{q})$ for the five bands forming a Fermi surface, for several choices of the form factor functions, whose momentum dependence is displayed in the last row of Fig. S7. We observe that for most form factor channels the $\alpha$ band has the largest susceptibility with peaks at the M and L points. In the following, we will focus on the $\alpha$ band because of its large contribution to the susceptibilities peaked at M and L points so that it likely plays an important role in the formation of a CDW as a Fermi-surface instability. We note that there are competing inter-band processes with wave vector $K$. However, their analysis is beyond our first approach here where we focus on intra-band effects, which is also motivated by the experimental observation of wave vectors $M$ and $L$.

## SII. 2D AND SINGLE BAND APPROXIMATION

As shown in Fig. S4, we expect the $\alpha$ band to be the most important one for the system's correlations. For this reason, we isolate and look at this single band in the following analysis. Furthermore, we would like to make use of the quasi-two-dimensional behavior of its band susceptibility. Our goal is to analyze the leading Fermi-surface instabilities of the system, including the competition of all interaction channels, via an unbiased functional renormalization group

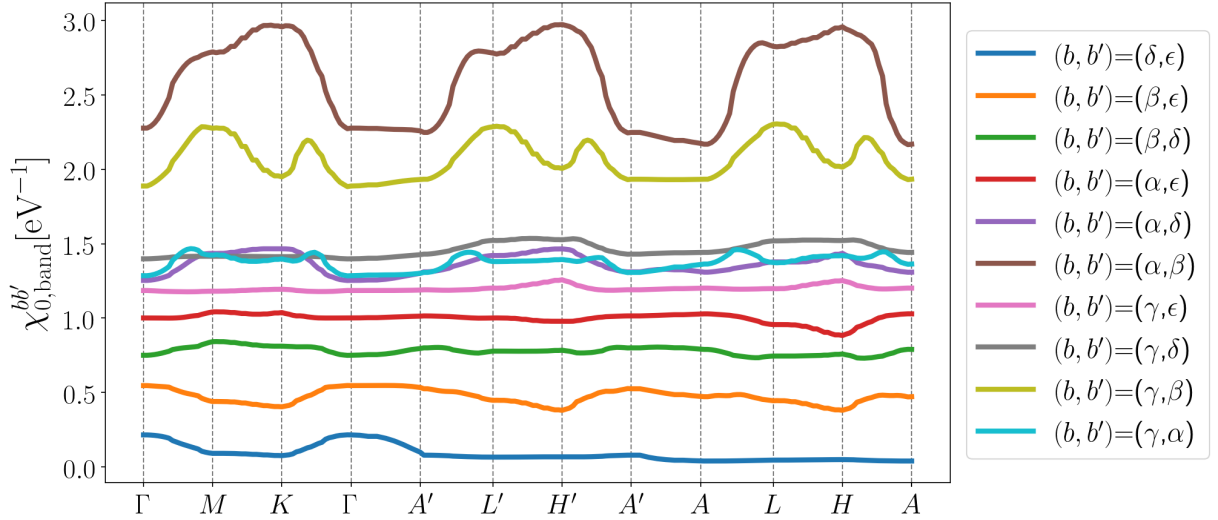

Figure S5. Interband susceptibility [see Eq. (S1)]. As in Fig. S4, all calculations have been performed at $T = 30$ meV. The points used for the Brillouin zone path are displayed in the right panel of Fig. S4.

(fRG) approach. However, performing an fRG calculation on the three-dimensional band structure with multiple Fermi surfaces is numerically extremely expensive (see Sec. SIV).

To reduce the system's dimensionality, we determine the two-dimensional cuts that give the largest contribution to in the integral over $k_z$ defining the three-dimensional band susceptibility $\chi^{\alpha}_{0,band}$ (Eq. (S1)). To this end, we consider two dimensional cuts at fixed $k_z$ where the Fermi surfaces exhibit two-dimensional-like van Hove singularities, because these likely give the dominant contribution to $\chi^{\alpha}_{0,band}$. These are points where $\epsilon^{\ell}_{\mathbf{k}} \sim \alpha k_x^2 - \beta k_y^2$ for $\mathbf{k}$ close to one of the $M$ points, with $\epsilon^{\ell}_{\mathbf{k}}$ the band dispersion and $\alpha, \beta > 0$ some coefficients. Note, however, that if the cut is performed at $k_z \notin \{0, \pi\}$, a term linear in $k_z$ will also be present. These cuts are made exactly at the $k_z$ values marked with gray dotted lines in Fig. S3, where the two dimensional cuts of the $\alpha$-sheet exhibit a change in their two-dimensional topology. In panel (a) of Fig. S8, we show the three $k_z$-cuts where the Fermi surfaces exhibit a two-dimensional-like van Hove singularity. Note that for each cut at fixed $k_z$, we have an identical Fermi surface topology at $-k_z$. In order to quantify up to which level of accuracy a two-dimensional cut can reproduce the three-dimensional band susceptibility, we define the quantity

$$\chi^{\alpha}_{0,2D,i}(\mathbf{q}) = \int_{\text{2D BZ}} \frac{d^2\mathbf{k}}{(2\pi)^2} \frac{f(E^{\alpha}_{\mathbf{k+q},k_{z,i}}) - f(E^{\alpha}_{\mathbf{k},k_{z,i}})}{E^{\alpha}_{\mathbf{k},k_{z,i}} - E^{\alpha}_{\mathbf{k+q},k_{z,i}}} \tag{S8}$$

where 2D BZ indicated the two-dimensional Brillouin zone, and $k_{z,i}$ ($i = 0, 1, 2$) are the three values of $k_z$ at which the two-dimensional cut of the $\alpha$ sheet exhibits a change in topology. In Fig. S8b, we compare $\chi^{\alpha}_{0,\text{band}}$ with the three different $\chi^{\alpha}_{0,2D,i}$. We observe that the peak structure of $\chi^{\alpha}_{0,\text{band}}$ around the $L$-point is qualitatively well reproduced by the two-dimensional band susceptibility $\chi^{\alpha}_{0,2D,i}$ obtained at $k_z = k_{z,0} \sim 1.425$.

Having established that the $\alpha$-band is the one in which correlation effects are expected to be the strongest, and that its two-dimensional cut at $k_z \sim 1.425$ well reproduces the momentum dependence of the three-dimensional band susceptibility, we map our multi-band three-dimensional problem onto a single-band two-dimensional problem with band dispersion

$$\epsilon_{\mathbf{k}} = E^{\alpha}_{\mathbf{k},k_{z,0}}. \tag{S9}$$

In Fig. S9 we show the dispersion of $\epsilon_{\mathbf{k}}$ in the 2D BZ. We note that this approximation is expected to be qualitatively correct as long as states in the vicinity of the Fermi surface play the dominant role for correlations. As we will see below, our results indicate small energy scales connected to interaction-induced instabilities.

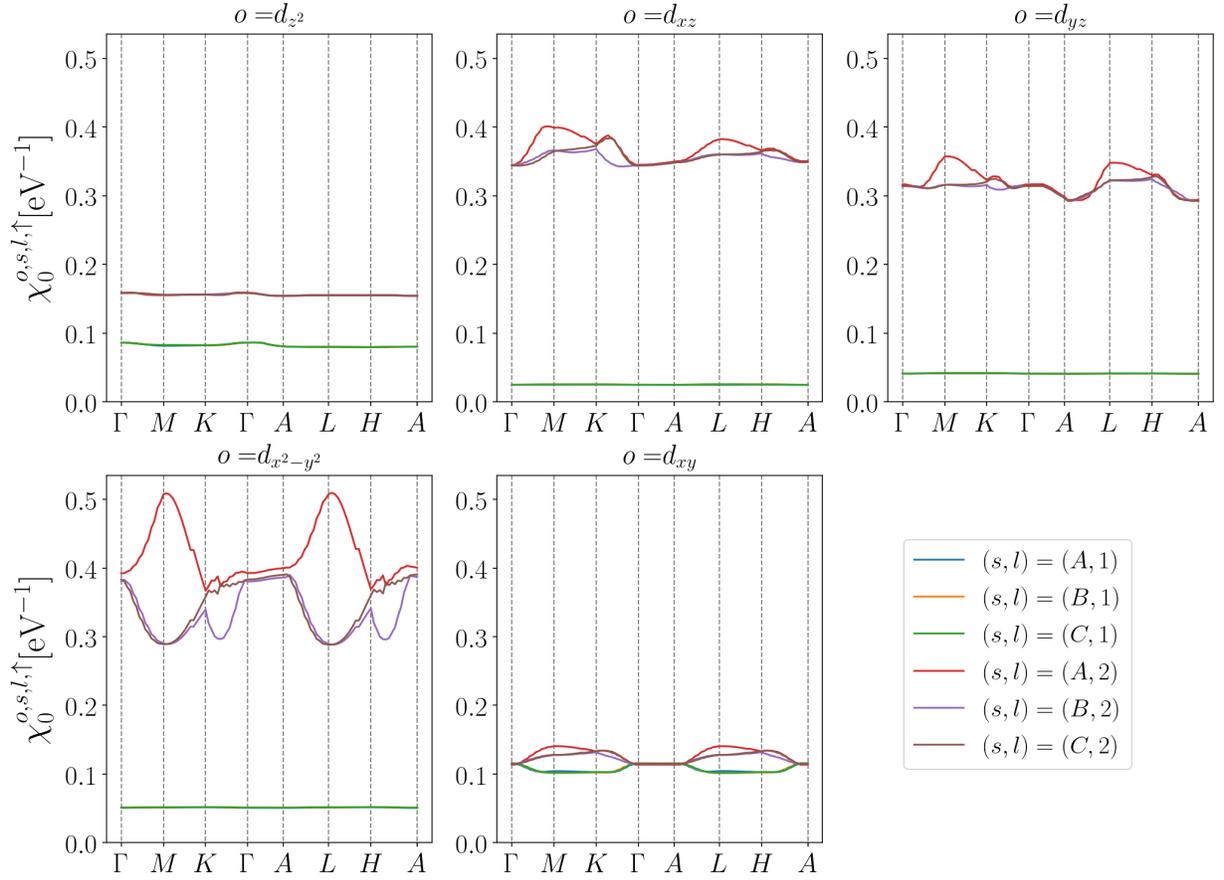

Figure S6. Orbital susceptibilities $\chi_0^{o,s,l,\uparrow}(\mathbf{q})$ for the Fe $d$-orbitals.

|  | $U$ | $U'_{\uparrow\downarrow}$ | $U'_{\uparrow\uparrow,1}$ | $U'_{\uparrow\uparrow,2}$ | $J_{\uparrow,1}$ | $J_{\uparrow,2}$ | $U_1$ |
|---|---|---|---|---|---|---|---|
| [eV] | 5.72 | 3.92 | 3.96 | 3.88 | 0.92 | 0.88 | 2.30 |

Table SI. Interaction parameters as computed from cRPA.

## SIII. PROJECTION ONTO THE FERMI SURFACE

We now turn our attention onto the interacting term of our effective Hamiltonian. In the antiferromagnetic phase (characterized by the spin-dependent $U'$ and $J$ couplings), the Hubbard-Kanamori Hamiltonian takes the form

$$\begin{aligned}\mathcal{H}_{\text{int}} =& U\sum_{j,l,o} n_{j,l,o,\uparrow} n_{j,l,o,\downarrow} + \frac{1}{2}\sum_{j,l,o\neq o',\sigma} U'_{\sigma\bar{\sigma},l}\, n_{j,l,o,\sigma}\, n_{j,l,o',\bar{\sigma}} + \frac{1}{2}\sum_{j,o\neq o',\sigma} (U'_{\sigma\sigma,l} - J_{\sigma,l})\, n_{j,l,o,\sigma}\, n_{j,l,o',\sigma} \\ & - \frac{1}{2}\sum_{j,l,o\neq o',\sigma} J_{\sigma,l}\, c^\dagger_{j,o,l,\sigma} c_{j,o,l,\bar\sigma} c^\dagger_{j,o',l,\bar\sigma} c_{j,o',l,\sigma} + \frac{1}{2}\sum_{j,l,o\neq o',\sigma} J_{\sigma,l}\, c^\dagger_{j,o,l,\sigma} c^\dagger_{j,o,l,\bar\sigma} c_{j,o',l,\bar\sigma} c_{j,o',l,\sigma} \\ & + \frac{1}{2} U_1 \sum_{\substack{\langle j,j'\rangle,l \\ o,o',\sigma,\sigma'}} n_{j,l,o,\sigma}\, n_{j',l,o',\sigma'}, \end{aligned} \quad (S10)$$

where $o \in \{d_{z^2}, d_{xz}, d_{yz}, d_{x^2-y^2}, d_{xy}\}$ runs over the Fe $d$-orbitals, $l \in \{1,2\}$ labels the layers, $\sigma \in \{\uparrow,\downarrow\}$ the spin projection along the $z$-axis, and $n_{j,l,o,\sigma} = c^\dagger_{j,o,l,\sigma} c_{j,o,l,\sigma}$. We have also introduced the convention $\bar\uparrow = \downarrow$, and $\bar\downarrow = \uparrow$. The operator $c_{j,o,l,\sigma}$ ($c^\dagger_{j,o,l,\sigma}$) annihilates (creates) an electron at lattice site $j$, with layer index $l$, orbital index $m$ and spin index $\sigma$. The antiferromagnetic state possesses a discrete symmetry implemented by the operator $T_z \otimes \mathcal{T}$, where $T_z$ is a one-site translation along the $c$-axis, and $\mathcal{T}$ is the time-reversal symmetry operator. This symmetry ensures that all the bands discussed in Sec. SI are doubly degenerate, due to a generalization of Kramers' theorem. This means that

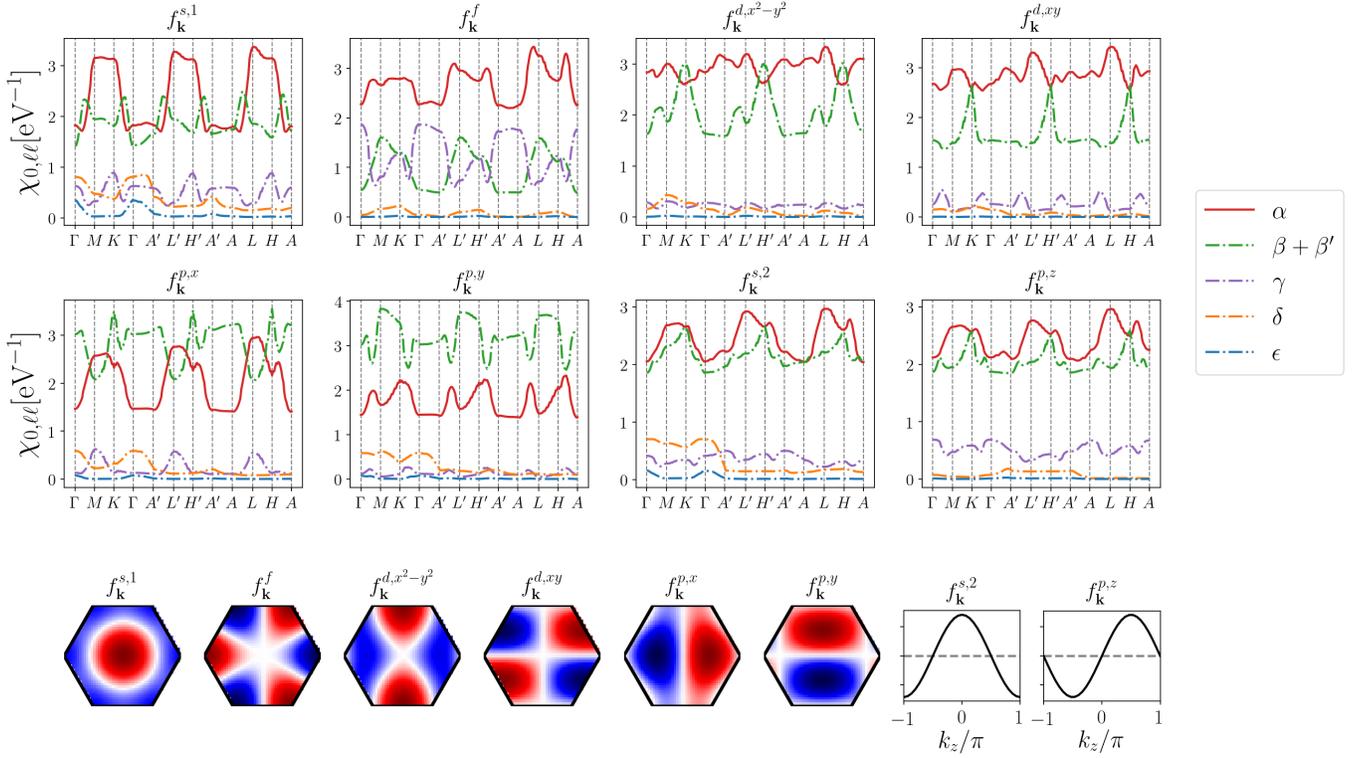

Figure S7. Diagonal ($\ell = \ell'$) form factor projected band susceptibility [Eq. (S7)] for each of the bands forming a Fermi surface. Here, as form factors we selected some low order lattice harmonics, whose momentum dependence is shown in the last row. Note that the first six functions depend exclusively on the in-plane momentum ($k_x, k_y$), while the remaining two only on the out-of-plane momentum $k_z$.

spin and layer are locked and we can label this degeneracy either with the layer index $l$ or with the spin projection $\sigma$. In the following we will choose the latter. Moreover, this symmetry implies the following relations for the interaction parameters

$$U'_{\uparrow\downarrow,l} = U'_{\downarrow\uparrow,l} \equiv U'_{\uparrow\downarrow}, \tag{S11a}$$

$$U'_{\uparrow\uparrow,l} = U'_{\downarrow\downarrow,\bar{l}}, \tag{S11b}$$

$$J_{\uparrow,l} = J_{\downarrow,\bar{l}}, \tag{S11c}$$

with $\bar{1} = 2$ and $\bar{2} = 1$. In Table SI, we report the computed values for the interaction parameters.

To project the interaction onto the $\alpha$-band, we approximate

$$c_{\mathbf{k},o,s,l,\sigma} \approx \phi_{\mathbf{k},o,s,l,\sigma} a_{\mathbf{k},\sigma} \tag{S12}$$

where $c_{\mathbf{k},o,s,l,\sigma}$ is the Fourier transform of $c_{j,o,l,\sigma}$, the index $s = \in \{A, B, C\}$ labels the three inequivalent Fe atoms in the (nonmagnetic) unit cell, and $a_{\mathbf{k},\sigma}$ annihilates an electron in the $\alpha$-band with momentum $\mathbf{k}$ and spin projection $\sigma$. The function $\phi_{\mathbf{k},o,s,l,\sigma}$ is the eigenvector of the *ab-initio* Hamiltonian and, due to the discrete symmetry of the AF state, it obeys the relation

$$\phi_{\mathbf{k},o,s,l,\bar{\sigma}} = -\text{sign}(\sigma) \phi_{\mathbf{k},o,s,\bar{l},\sigma}, \tag{S13}$$

with $\bar{1} = 2$, $\bar{2} = 1$, $\bar{\uparrow} = \downarrow$, $\bar{\downarrow} = \uparrow$, $\text{sign}(\uparrow) = +1$, and $\text{sign}(\downarrow) = -1$.

The eigenvector $\phi_{\mathbf{k},o,s,l,\sigma}$ is not uniquely defined. Indeed, one can multiply it by an arbitrary $\mathbf{k}$ dependent phase factor $e^{i\varphi_\mathbf{k}}$ and it will remain an eigenvector of the ab-initio Hamiltonian. Because they contain the eigenvectors calculated at different momenta, the projected interactions will depend on the specific choice of $\varphi_\mathbf{k}$. We overcome this difficulty by enforcing that $\phi_{\mathbf{k},\tilde{o},\tilde{s},\tilde{m},\uparrow}$ is real, where $\tilde{o}, \tilde{s}, \tilde{m}$ are the values of $b, l, m$ that maximize $|\phi_{\mathbf{k},o,s,l,\uparrow}|$. In this way we are able to obtain a function $\phi_{\mathbf{k},o,s,l,\sigma}$ that is continuous as a function of $\mathbf{k}$ and has a in general very small

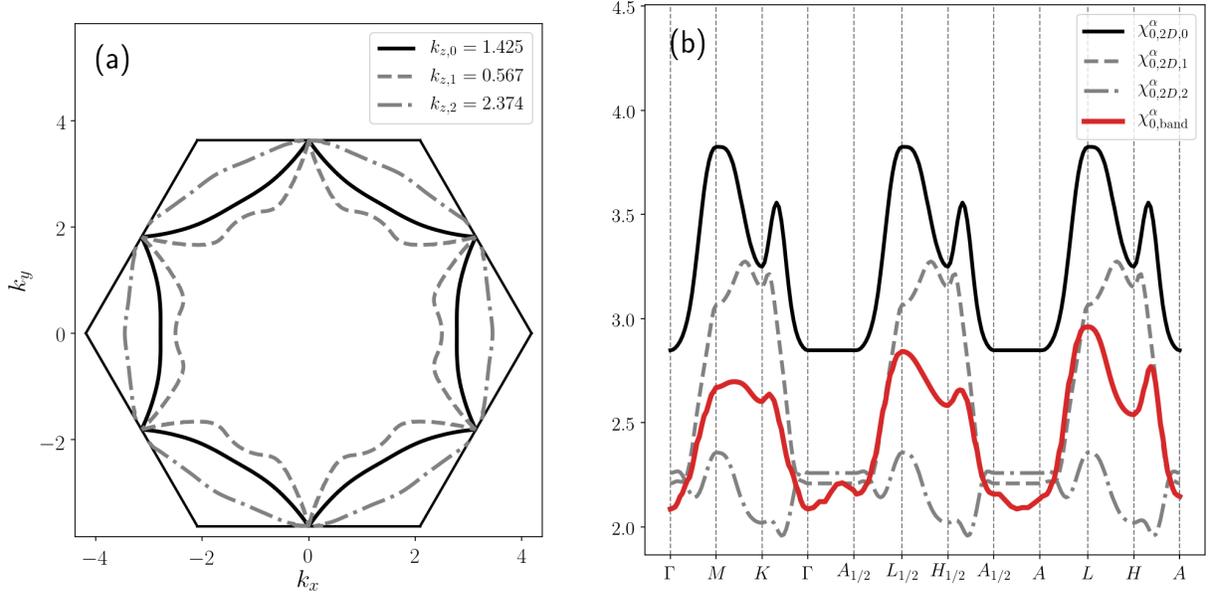

Figure S8. Panel (a): two-dimensional cuts of the Fermi surface $\alpha$-sheet at which a change in topology occurs. Panel (b): comparison of the three-dimensional band susceptibility $\chi^\alpha_{0,\text{band}}$ (red curve) of the Fermi surface $\alpha$ sheet with three two-dimensional band susceptibilities $\chi^\alpha_{0,2D,i}$ (black and gray curves, see Eq. (S8)) obtained at the three different two-dimensional cuts in panel (a). The susceptibilities in panel (b) have been computed at $T = 30$ meV. The units on the $y$-axis of panel (b) are eV$^{-1}$

imaginary part. Moreover, we require $\phi_{\mathbf{k},o,s,l,\uparrow}$ to transform under $C_6$ rotations as

$$\phi_{\mathbf{k},o,s,l,\sigma} \to \sum_{s' \in \{A,B,C\}} U_{s,s'}\, \phi_{C_6\mathbf{k},o,s',l,\sigma}\,, \tag{S14}$$

where $U = \begin{pmatrix} 0 & 0 & 1 \\ 1 & 0 & 0 \\ 0 & 1 & 0 \end{pmatrix}$ is a $3 \times 3$ matrix that rotates between the three Fe-kagome sublattices, and $C_6\mathbf{k}$ is the $\mathbf{k}$-point rotated by 60 degrees clockwise.

Inserting (S12) into (S10), we get

$$\mathcal{H}_{\text{int,proj}} = \int_{\substack{\mathbf{k}_1,\mathbf{k}_2 \\ \mathbf{k}_3,\mathbf{k}_4}} \sum_{\substack{\sigma_1,\sigma_2 \\ \sigma_3,\sigma_4}} V_{\sigma_1\sigma_2\sigma_3\sigma_4}(\mathbf{k}_1,\mathbf{k}_2,\mathbf{k}_3,\mathbf{k}_4)\, a^\dagger_{\mathbf{k}_1,\sigma_1} a^\dagger_{\mathbf{k}_2,\sigma_2} a_{\mathbf{k}_4,\sigma_4} a_{\mathbf{k}_3,\sigma_3}\, \delta_{\mathbf{k}_4,\mathbf{k}_1+\mathbf{k}_2-\mathbf{k}_3}\,, \tag{S15}$$

with $\int_{\mathbf{k}} = \int_{\text{BZ}} \frac{d^3\mathbf{k}}{(2\pi)^3}$. Exploiting the discrete symmetry mentioned above, as well as the crossing symmetries of (S15), we can reduce the number of independent components of $V_{\sigma_1\sigma_2\sigma_3\sigma_4}(\mathbf{k}_1,\mathbf{k}_2,\mathbf{k}_3,\mathbf{k}_4)$ to two:

$$V_{\uparrow\uparrow\uparrow\uparrow}(\mathbf{k}_1,\mathbf{k}_2,\mathbf{k}_3,\mathbf{k}_4) = V_{\downarrow\downarrow\downarrow\downarrow}(\mathbf{k}_1,\mathbf{k}_2,\mathbf{k}_3,\mathbf{k}_4) \equiv V_{\uparrow\uparrow}(\mathbf{k}_1,\mathbf{k}_2,\mathbf{k}_3,\mathbf{k}_4)\,, \tag{S16a}$$

$$V_{\uparrow\downarrow\uparrow\downarrow}(\mathbf{k}_1,\mathbf{k}_2,\mathbf{k}_3,\mathbf{k}_4) = V_{\downarrow\uparrow\downarrow\uparrow}(\mathbf{k}_1,\mathbf{k}_2,\mathbf{k}_3,\mathbf{k}_4) \equiv V_{\uparrow\downarrow}(\mathbf{k}_1,\mathbf{k}_2,\mathbf{k}_3,\mathbf{k}_4)\,, \tag{S16b}$$

$$V_{\uparrow\downarrow\downarrow\uparrow}(\mathbf{k}_1,\mathbf{k}_2,\mathbf{k}_3,\mathbf{k}_4) = V_{\downarrow\uparrow\uparrow\downarrow}(\mathbf{k}_1,\mathbf{k}_2,\mathbf{k}_3,\mathbf{k}_4) \equiv V_{\overline{\uparrow\downarrow}}(\mathbf{k}_1,\mathbf{k}_2,\mathbf{k}_3,\mathbf{k}_4) = -V_{\uparrow\downarrow}(\mathbf{k}_2,\mathbf{k}_1,\mathbf{k}_3,\mathbf{k}_4)\,. \tag{S16c}$$

$V_{\uparrow\uparrow}$ and $V_{\uparrow\downarrow}$ are given by

$$V_{\uparrow\uparrow}(\mathbf{k}_1,\mathbf{k}_2,\mathbf{k}_3,\mathbf{k}_4) = \frac{1}{4}\big(V^{\text{NS}}_{\uparrow\uparrow}(\mathbf{k}_1,\mathbf{k}_2,\mathbf{k}_3,\mathbf{k}_4) - V^{\text{NS}}_{\uparrow\uparrow}(\mathbf{k}_2,\mathbf{k}_1,\mathbf{k}_3,\mathbf{k}_4)$$
$$+ V^{\text{NS}}_{\uparrow\uparrow}(\mathbf{k}_2,\mathbf{k}_1,\mathbf{k}_4,\mathbf{k}_3) - V^{\text{NS}}_{\uparrow\uparrow}(\mathbf{k}_1,\mathbf{k}_2,\mathbf{k}_4,\mathbf{k}_3)\big)\,, \tag{S17a}$$

$$V_{\uparrow\downarrow}(\mathbf{k}_1,\mathbf{k}_2,\mathbf{k}_3,\mathbf{k}_4) = \frac{1}{2}\big(V^{\text{NS}}_{\uparrow\downarrow}(\mathbf{k}_1,\mathbf{k}_2,\mathbf{k}_3,\mathbf{k}_4) + V^{\text{NS}}_{\uparrow\downarrow}(\mathbf{k}_2,\mathbf{k}_1,\mathbf{k}_4,\mathbf{k}_3)\big)\,, \tag{S17b}$$


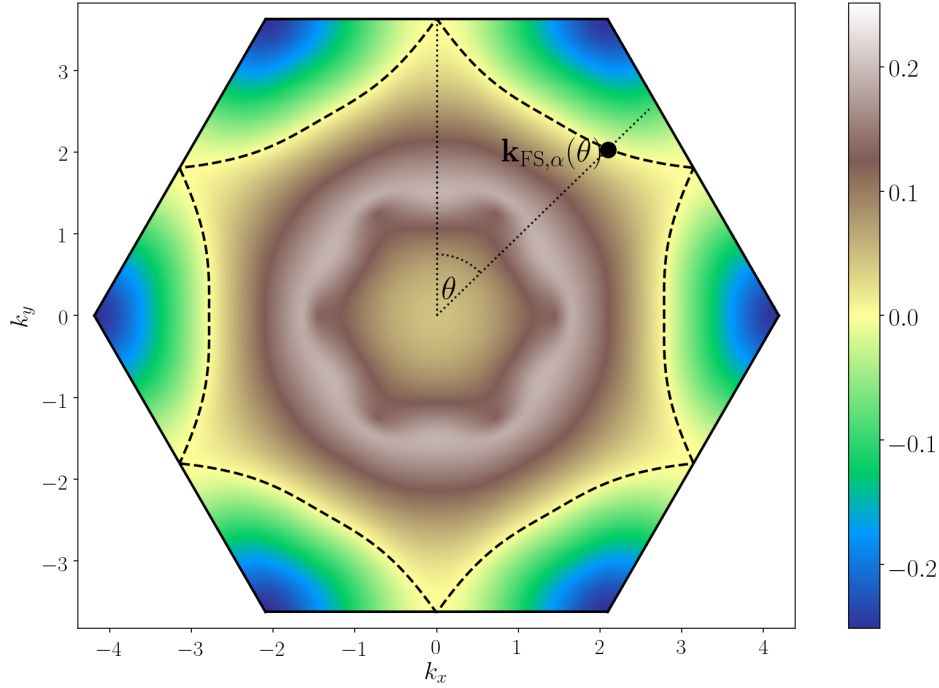

Figure S9. Dispersion of $\epsilon_\mathbf{k}$ in the 2D BZ [see Eq. (S9)]. The dashed line indicates the 2D Fermi surface. $\mathbf{k}_{\mathrm{FS},\alpha}(\theta)$ denotes the momentum on the Fermi surface lying on a straight line departing from the $\Gamma$ point ($\mathbf{k} = (0,0)$) and forming an angle $\theta$ with the $k_y$ axis. The colorbar units are eV.

with

$$V^{\mathrm{NS}}_{\uparrow\uparrow}(\mathbf{k}_1, \mathbf{k}_2, \mathbf{k}_3, \mathbf{k}_4) = \frac{1}{2} \sum_{\substack{s,l \\ o\neq o'}} (U'_{\uparrow\uparrow,l} - J_{\uparrow,l})\, \phi^*_{\mathbf{k}_1,o,s,l,\uparrow}\phi^*_{\mathbf{k}_2,o,s',l,\uparrow}\phi_{\mathbf{k}_3,o,s,l,\uparrow}\phi_{\mathbf{k}_4,o,s',l,\uparrow}$$
$$+ U_1 \sum_{\substack{s,s',l \\ o,o'}} v^{bb'}_{\mathbf{k}_3-\mathbf{k}_1}\, \phi^*_{\mathbf{k}_1,o,s,l,\uparrow}\phi^*_{\mathbf{k}_2,b',m',l,\uparrow}\phi_{\mathbf{k}_3,o,s,l,\uparrow}\phi_{\mathbf{k}_4,b',m',l,\uparrow}, \qquad (\mathrm{S18a})$$

$$V^{\mathrm{NS}}_{\uparrow\downarrow}(\mathbf{k}_1, \mathbf{k}_2, \mathbf{k}_3, \mathbf{k}_4) = U \sum_{s,l,o} \phi^*_{\mathbf{k}_1,o,s,l,\uparrow}\phi^*_{\mathbf{k}_2,o,s,l,\downarrow}\phi_{\mathbf{k}_3,o,s,l,\uparrow}\phi_{\mathbf{k}_4,o,s,l,\downarrow}$$
$$+ \frac{1}{2} U'_{\uparrow\downarrow} \sum_{\substack{s,l \\ o\neq o'}} \phi^*_{\mathbf{k}_1,o,s,l,\uparrow}\phi^*_{\mathbf{k}_2,o,s',l,\downarrow}\phi_{\mathbf{k}_3,o,s,l,\uparrow}\phi_{\mathbf{k}_4,o,s',l,\downarrow}$$
$$- \frac{1}{2} \sum_{\substack{s,l \\ o\neq o'}} J_{\uparrow,l}\, \phi^*_{\mathbf{k}_1,o,s,l,\uparrow}\phi^*_{\mathbf{k}_2,o,s',l,\downarrow}\phi_{\mathbf{k}_3,o,s,l,\downarrow}\phi_{\mathbf{k}_4,o,s',l,\uparrow}$$
$$+ \frac{1}{2} \sum_{\substack{s,l \\ o\neq o'}} J_{\uparrow,l}\, \phi^*_{\mathbf{k}_1,o,s,l,\uparrow}\phi^*_{\mathbf{k}_2,o,s,l,\downarrow}\phi_{\mathbf{k}_3,o,s',l,\downarrow}\phi_{\mathbf{k}_4,o,s',l,\uparrow}$$
$$+ U_1 \sum_{\substack{s,s',l \\ o,o'}} v^{ss'}_{\mathbf{k}_3-\mathbf{k}_1}\, \phi^*_{\mathbf{k}_1,o,s,l,\uparrow}\phi^*_{\mathbf{k}_2,b',m',l,\downarrow}\phi_{\mathbf{k}_3,o,s,l,\uparrow}\phi_{\mathbf{k}_4,b',m',l,\downarrow}. \qquad (\mathrm{S18b})$$

The matrix $v^{ss'}_\mathbf{q}$ is given by the Fourier transform of the nearest-neighbor interaction. Therefore, only its off-diagonal entries (coupling different kagome sublattices) are nonvanishing and dependent on the transfer momentum $\mathbf{q}$. It takes

the form

$$\boldsymbol{v_q} = \begin{pmatrix} 0 & \cos\left(\frac{q_x}{4} - \frac{\sqrt{3}q_y}{4}\right) & \cos\left(\frac{q_x}{4} + \frac{\sqrt{3}q_y}{4}\right) \\ \cos\left(\frac{q_x}{4} - \frac{\sqrt{3}q_y}{4}\right) & 0 & \cos\left(\frac{q_x}{2}\right) \\ \cos\left(\frac{q_x}{4} + \frac{\sqrt{3}q_y}{4}\right) & \cos\left(\frac{q_x}{2}\right) & 0 \end{pmatrix}. \tag{S19}$$

We can see the explicit dependence of the interactions on the eigenvector $\phi_{\mathbf{k},o,s,l,\sigma}$.

The eigenvector is not unique, e.g., an alternative choice for the matrix $\boldsymbol{v_q}$ can be obtained by multiplying with a phase factor that removes the sublattice shifts within a unit cell, resulting in

$$\boldsymbol{v_q} = \begin{pmatrix} 0 & 1 + e^{2i\left(\frac{q_x}{4} - \frac{\sqrt{3}q_y}{4}\right)} & 1 + e^{-2i\left(\frac{q_x}{4} + \frac{\sqrt{3}q_y}{4}\right)} \\ 1 + e^{-2i\cos\left(\frac{q_x}{4} - \frac{\sqrt{3}q_y}{4}\right)} & 0 & 1 + e^{2i\left(\frac{q_x}{2}\right)} \\ 1 + e^{2i\left(\frac{q_x}{4} + \frac{\sqrt{3}q_y}{4}\right)} & 1 + e^{-2i\left(\frac{q_x}{2}\right)} & 0 \end{pmatrix}. \tag{S20}$$

Here, we choose to include the sublattice shifts as in Eq. (S19), a choice which is consistent with the *ab initio* calculations. Note that one could also use the form (S20) for $\boldsymbol{v_q}$ and change basis in the *ab-initio* eigenvectors $\phi_{\mathbf{k},o,s,l,\sigma}^b$ by removing the sublattice shifts. The two choices are equivalent and should not modify the physical outcome of our analysis. We caution, however, that the approximation of projecting into a single band can lead to discrepancies between different choices. These are small as long as contributions from high-energy states away from the Fermi level are negligible [12].

In summary, the projected Hamiltonian restricted to the two-dimensional cut at $k_z = k_{z,0} \sim 1.425$ is given by

$$\mathcal{H}_{\text{proj,2D}} = \int_{\mathbf{k}} \sum_\sigma \epsilon_\mathbf{k}\, a^\dagger_{\mathbf{k},\sigma} a_{\mathbf{k},\sigma} \tag{S21a}$$

$$+ \int_{\substack{\mathbf{k}_1,\mathbf{k}_2 \\ \mathbf{k}_3,\mathbf{k}_4}} \sum_{\substack{\sigma_1,\sigma_2 \\ \sigma_3,\sigma_4}} V_{\sigma_1\sigma_2\sigma_3\sigma_4}(\mathbf{k}_1,\mathbf{k}_2,\mathbf{k}_3,\mathbf{k}_4)\, a^\dagger_{\mathbf{k}_1,\sigma_1} a^\dagger_{\mathbf{k}_2,\sigma_2} a_{\mathbf{k}_4,\sigma_4} a_{\mathbf{k}_3,\sigma_3}\, \delta_{\mathbf{k}_4,\mathbf{k}_1+\mathbf{k}_2-\mathbf{k}_3}, \tag{S21b}$$

where we have redefined $a_{\mathbf{k},\sigma} \to a_{\mathbf{k},k_{z,0},\sigma}$, $\int_{\mathbf{k}} \to \int_{\text{2D BZ}} \frac{d^2\mathbf{k}}{(2\pi)^2}$, and

$$V_{\sigma_1\sigma_2\sigma_3\sigma_4}(\mathbf{k}_1,\mathbf{k}_2,\mathbf{k}_3,\mathbf{k}_4) \to V_{\sigma_1\sigma_2\sigma_3\sigma_4}(\mathbf{k}_1,k_{z,0};\mathbf{k}_2,k_{z,0};\mathbf{k}_3,k_{z,0};\mathbf{k}_4,k_{z,0}).$$

### A. Sublattice interference mechanism and layer polarization

It is instructive to inspect the orbital content of the $\alpha$ Fermi surface sheet. With this goal in mind, we define the quantity

$$\varrho_{\theta,o,l} = \sum_{s=1}^{3} |\phi_{\mathbf{k}_{\text{FS}}(\theta),o,s,l,\uparrow}|^2, \tag{S22}$$

where $\mathbf{k}_{\text{FS}}(\theta)$ is the two-dimensional BZ point belonging to the cut at $k_{z,0}$ of the $\alpha$ Fermi surface sheet and forming an angle $\theta$ with the $k_y$ axis (see Fig. S9).

In Fig. S10a and S10c, we show the orbital content of the $\alpha$ Fermi surface sheet for $\sigma = \uparrow$ in layer 1 and 2, respectively. We observe that the largest weight occurs in layer 1. Within layer 1, the largest contribution comes from the $d_{xz}$ orbitals of Fe, with non negligible contributions from the Ge atoms and from the Fe $d_{z^2}$ orbitals. In Fig. S10b and S10d we show the kagome sublattice resolved weight for Fe $d_{xz}$ orbitals, $\varrho_{\theta,b,d_{xz},l} = |\phi_{\mathbf{k}_{\text{FS}}(\theta),b,d_{xz},l,\uparrow}|^2$, in layer 1 and 2, respectively. We note that at each $M$-point (indicated by vertical dotted gray lines) only one of the three sublattices has a nonvanishing weight. This realizes the so-called sublattice interference mechanism, which, in presence of purely local interactions, suppresses all $\mathbf{q} \neq 0$ instabilities [13]. In addition, due to the strong layer polarization, unequal spin scattering processes are also suppressed in our case.

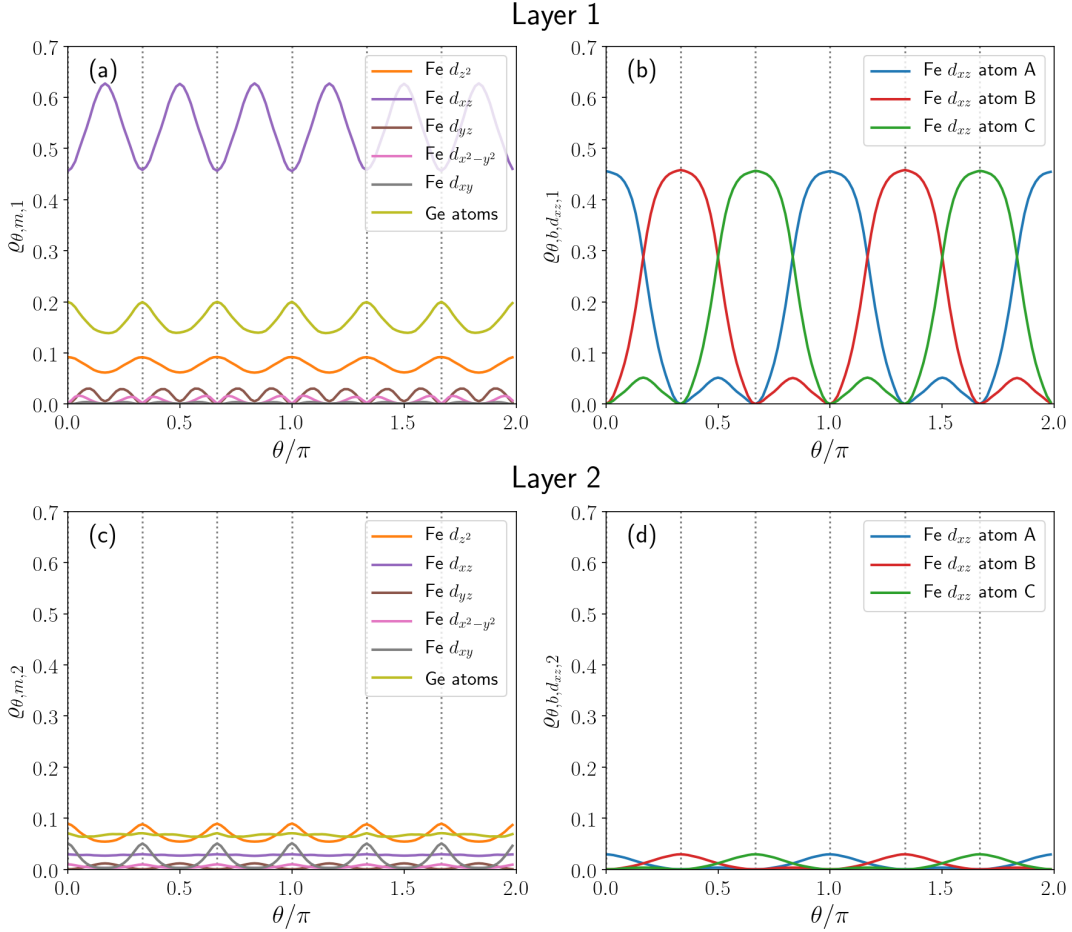

Figure S10. Orbital and layer content of the $\alpha$ Fermi sheet at $k_z = k_{z,0} \sim 1.452$. In panels (a) and (c) we show the orbital content in layers 1 and 2, respectively. In panels (b) and (c) we plot the sublattice resolved orbital content for the Fe $d_{xz}$ orbitals in layers 1 and 2, respectively.

## SIV. PATCH FRG

The fRG approach to correlated electron systems has been established as a versatile method to identify the leading Fermi-surface instabilities without introducing a bias towards a specific mean-field channel. Schematically, the fRG can be seen as a functional version of Wilsonian RG, which is initialized at an ultraviolet scale $\Lambda_{\mathrm{UV}}$ and successively integrates out (fermionic) fluctuations while the scale is lowered towards the infrared $\Lambda_{\mathrm{IR}} \to 0$. While the fRG framework generally implies a scale-dependence of correlation functions of all orders, a concrete numerical evaluation of the corresponding RG flow equations typically requires truncations, that is, six-point and higher correlation functions are neglected and set to zero [14]. Our present implementation for electronic systems with competing interactions, is focused on the evolution of the two-particle interaction vertex to detect the leading Fermi-surface instabilities, i.e., we also neglect self-energy effects.

The fRG approach is based on the functional integral formalism for many-electron systems with action

$$S[\overline{\psi}, \psi] = -(\overline{\psi}, G_0^{-1}\psi) + S_{\mathrm{int}}[\overline{\psi}, \psi], \tag{S23}$$

where $\overline{\psi}, \psi$ are Grassmann fields, $G_0(\omega, \boldsymbol{k}) = 1/(i\omega - \epsilon(\boldsymbol{k}))$ is the bare propagator with Matsubara frequency $\omega$, single-particle dispersion $\epsilon(\boldsymbol{k})$, and $(.,.)$ denotes integration over continuous and summation over discrete quantum numbers of the system, e.g., Matsubara frequency, momentum, and bands $(\overline{\psi}, \eta) = \sum_{i\omega} \int_{\boldsymbol{q}} \sum_b \psi_{\omega,\boldsymbol{q},b} \eta_{\omega,\boldsymbol{q},b}$. The formalism can be readily generalized to the case of propagators with several band, orbital, sublattice, layer, and spin indices [14, 15]. $S_{\mathrm{int}}[\overline{\psi}, \psi]$ is an interaction term that follows from the microscopic Hamiltonian of the system.

The action $S$ defines the the Schwinger functional $\mathcal{G}[\overline{\eta}, \eta] = -\ln \int D\psi D\overline{\psi} \exp(-S[\overline{\psi}, \psi]) \exp[(\overline{\eta}, \psi) + (\overline{\psi}, \eta)]$ and by Legendre transformation, we obtain the effective action $\Gamma[\overline{\psi}, \psi] = (\overline{\eta}, \psi) + (\overline{\psi}, \eta) + G[\overline{\eta}, \eta]$ with $\psi = -\partial G/\partial \overline{\eta}$ and

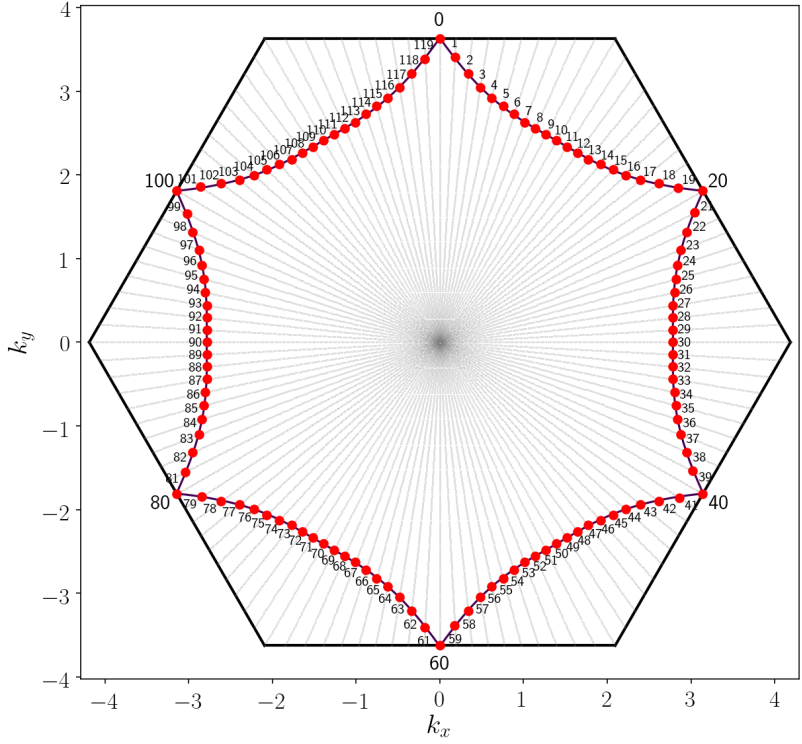

Figure S11. Fermi surface patches used for the fRG computations.

$\overline{\psi} = \partial G/\partial \eta$, which generates the one-particle irreducible correlation functions.

To obtain functional renormalization group flow equations, we first regularize the bare propagator by an infrared cutoff $\Lambda$, i.e., we introduce the modification $G_0(\omega, \boldsymbol{k}) \to G_0^\Lambda(\omega, \boldsymbol{k})$. There are different ways to implement the regularisation. We use the temperature cutoff introduced in Ref. [16]. The modified propagator now induces a scale dependence of the effective action, which is described by the exact renormalization group flow equation

$$\frac{\partial}{\partial \Lambda}\Gamma^\Lambda = -(\overline{\psi}, (\dot{G}_0^\Lambda)^{-1}\psi) - \frac{1}{2}\mathrm{Tr}\left((\dot{\mathbf{G}}_0^\Lambda)^{-1}(\mathbf{\Gamma}^{(2)\Lambda})^{-1}\right), \tag{S24}$$

where $\mathbf{\Gamma}^{(2)\Lambda} = (\partial_{\overline{\psi}}, \partial_\psi)^T(\partial_\psi, \partial_{\overline{\psi}})\Gamma^\Lambda$ denotes the second derivatives of $\Gamma^\Lambda$. We here introduce a temperature cutoff scheme [16]. Coming back to the issue of truncation and approximations for numerical feasibility, we note that we here use a standard scheme, which (1) neglects self-energy corrections, (2) sets external Matsubara frequencies to zero, i.e., neglects the frequency dependence of the two-particle interaction vertex, and (3) truncates the three- and all higher-particle vertices. This scheme includes all leading contributions to instabilities coming from the Fermi surface. Eventually, we obtain fRG flow equations for the static two-particle interaction vertex $V_{\sigma\sigma'}(\boldsymbol{k}_1, \boldsymbol{k}_2, \boldsymbol{k}_3)$, schematically reading

$$\partial_\Lambda V_{\sigma\sigma'}(\boldsymbol{k}_1, \boldsymbol{k}_2, \boldsymbol{k}_3) = \beta_{V_{\sigma,\sigma'}}, \tag{S25}$$

We will discuss the concrete form of $\beta_{V_{\sigma,\sigma'}}$ for our case in Eqs. (S27),(S28) below. We can then determine the leading Fermi-surface instabilities from the singular contributions to $V_{\sigma\sigma'}$ that appear upon integrating out the fRG flow equations towards $\Lambda = 0$. As $V_{\sigma\sigma'}$ is a function of three independent wave vectors the evaluation is numerically costly, even in the case of a single band.

To simplify the wave vector dependence one can employ the so-called $N$-patch scheme where one approximates the wave vector dependence of the two-particle vertex as constant along elongated patches in the Brillouin zone, i.e., one projects onto the Fermi surface. We have chosen the $N$-patch scheme for consistency reasons with our quasi 2D, single-band approximation. For the concrete implementation we introduce an angular discretization of the Fermi surface, represented by $N$ patch points, see Fig. S11. Wave vectors $\boldsymbol{k}$ in the Brillouin zone are then mapped to their closest patch, i.e., the interaction vertex is fully determined by its value on the patch points. We note, that this scheme incorporates wave vector conservation only approximately, because the fourth wave vector of the projected vertex

will, in general, only lie on a patch point after additional projection. This strategy was still successfully employed to identify the leading Fermi-surface instabilities of many correlated electron systems, see Refs. [14, 15] for reviews. In conclusion, we obtain flow equations for two vertices with three projected momenta each, i.e., for $N$ patches, the numerical cost scales with $N^3$. Here, we implemented a Fermi-surface resolution of $N = 120$.

The initial interactions, after projection onto the Fermi surface, read

$$W_{\text{ini}}(\theta_1, \theta_2, \theta_3) = W(\mathbf{k}_{\text{FS}}(\theta_1), \mathbf{k}_{\text{FS}}(\theta_2), \mathbf{k}_{\text{FS}}(\theta_3), \text{proj}_{\text{FS}}[\mathbf{k}_{\text{FS}}(\theta_1) + \mathbf{k}_{\text{FS}}(\theta_2) - \mathbf{k}_{\text{FS}}(\theta_3)]) \,, \quad (S26)$$

with $W = V_{\uparrow\uparrow}$ or $V_{\uparrow\downarrow}$, and where the function $\text{proj}_{\text{FS}}[\mathbf{k}]$ returns the Fermi surface momentum that is closest to $\mathbf{k}$, modulo a reciprocal lattice vector. Following Ref. [16], we choose the temperature as our flow parameter $\Lambda \to T$. The flow equations for the Fermi-surface projected interactions read

$$\begin{aligned}
\partial_T V_{\uparrow\uparrow}(\theta_1, \theta_2, \theta_3) = &-\frac{1}{2}\frac{1}{N_\theta} \sum_{\bar{\theta}} V_{\uparrow\uparrow}(\theta_1, \theta_2, \bar{\theta}) \, L_{pp}(\bar{\theta}, \theta_1, \theta_2) \, V_{\uparrow\uparrow}(\bar{\theta}, \Theta_{\text{FS}}[\theta_1, \theta_2, \bar{\theta}], \theta_3) \\
&+ \frac{1}{N_\theta} \sum_{\bar{\theta}} V_{\uparrow\uparrow}(\theta_1, \bar{\theta}, \theta_3) \, L_{ph}(\bar{\theta}, \theta_1, \theta_3) \, V_{\uparrow\uparrow}(\Theta_{\text{FS}}[\bar{\theta}, \theta_3, \theta_1], \theta_2, \bar{\theta}) \\
&+ \frac{1}{N_\theta} \sum_{\bar{\theta}} V_{\uparrow\downarrow}(\theta_1, \bar{\theta}, \theta_3) \, L_{ph}(\bar{\theta}, \theta_1, \theta_3) \, V_{\uparrow\downarrow}(\Theta_{\text{FS}}[\bar{\theta}, \theta_3, \theta_1], \theta_2, \bar{\theta}) \\
&- \frac{1}{N_\theta} \sum_{\bar{\theta}} V_{\uparrow\uparrow}(\theta_2, \bar{\theta}, \theta_3) \, L_{ph}(\bar{\theta}, \theta_2, \theta_3) \, V_{\uparrow\uparrow}(\Theta_{\text{FS}}[\bar{\theta}, \theta_3, \theta_2], \theta_1, \bar{\theta}) \\
&- \frac{1}{N_\theta} \sum_{\bar{\theta}} V_{\uparrow\downarrow}(\theta_2, \bar{\theta}, \theta_3) \, L_{ph}(\bar{\theta}, \theta_2, \theta_3) \, V_{\uparrow\downarrow}(\Theta_{\text{FS}}[\bar{\theta}, \theta_3, \theta_2], \theta_1, \bar{\theta}) \,,
\end{aligned} \quad (S27)$$

$$\begin{aligned}
\partial_T V_{\uparrow\downarrow}(\theta_1, \theta_2, \theta_3) = &-\frac{1}{N_\theta} \sum_{\bar{\theta}} V_{\uparrow\downarrow}(\theta_1, \theta_2, \bar{\theta}) \, L_{pp}(\bar{\theta}, \theta_1, \theta_2) \, V_{\uparrow\downarrow}(\bar{\theta}, \Theta_{\text{FS}}[\theta_1, \theta_2, \bar{\theta}], \theta_3) \\
&+ \frac{1}{N_\theta} \sum_{\bar{\theta}} V_{\uparrow\downarrow}(\theta_1, \bar{\theta}, \theta_3) \, L_{ph}(\bar{\theta}, \theta_1, \theta_3) \, V_{\uparrow\uparrow}(\Theta_{\text{FS}}[\bar{\theta}, \theta_3, \theta_1], \theta_2, \bar{\theta}) \\
&+ \frac{1}{N_\theta} \sum_{\bar{\theta}} V_{\uparrow\uparrow}(\theta_1, \bar{\theta}, \theta_3) \, L_{ph}(\bar{\theta}, \theta_1, \theta_3) \, V_{\uparrow\downarrow}(\Theta_{\text{FS}}[\bar{\theta}, \theta_3, \theta_1], \theta_2, \bar{\theta}) \\
&- \frac{1}{N_\theta} \sum_{\bar{\theta}} V_{\uparrow\downarrow}(\theta_2, \bar{\theta}, \theta_3) \, L_{ph}(\bar{\theta}, \theta_2, \theta_3) \, V_{\uparrow\downarrow}(\Theta_{\text{FS}}[\bar{\theta}, \theta_3, \theta_2], \theta_1, \bar{\theta}) \,,
\end{aligned} \quad (S28)$$

where where $\Theta_{\text{FS}}[\theta_1, \theta_2, \theta_3]$ is a shorthand for $\text{proj}_{\text{FS}}[\mathbf{k}(\theta_1) + \mathbf{k}(\theta_2) - \mathbf{k}(\theta_3)]$, and we have defined the loop functions as

$$L_{pp}(\bar{\theta}, \theta_1, \theta_2) = \frac{\sqrt{3}}{4\pi} \int_0^{r_{\max}(\bar{\theta})} dr \, \frac{\partial_T f\left(\epsilon_{r(\cos\bar{\theta}, \sin\bar{\theta})}\right) + \partial_T f\left(\epsilon_{\mathbf{k}_{\text{FS}}(\theta_1) + \mathbf{k}_{\text{FS}}(\theta_2) - r(\cos\bar{\theta}, \sin\bar{\theta})}\right)}{\epsilon_{r(\cos\bar{\theta}, \sin\bar{\theta})} + \epsilon_{\mathbf{k}_{\text{FS}}(\theta_1) + \mathbf{k}_{\text{FS}}(\theta_2) - r(\cos\bar{\theta}, \sin\bar{\theta})}} \,, \quad (S29a)$$

$$L_{ph}(\bar{\theta}, \theta_1, \theta_3) = -\frac{\sqrt{3}}{4\pi} \int_0^{r_{\max}(\bar{\theta})} dr \, \frac{\partial_T f\left(\epsilon_{r(\cos\bar{\theta}, \sin\bar{\theta})}\right) - \partial_T f\left(\epsilon_{\mathbf{k}_{\text{FS}}(\theta_1) + r(\cos\bar{\theta}, \sin\bar{\theta}) - \mathbf{k}_{\text{FS}}(\theta_3)}\right)}{\epsilon_{r(\cos\bar{\theta}, \sin\bar{\theta})} - \epsilon_{\mathbf{k}_{\text{FS}}(\theta_1) + r(\cos\bar{\theta}, \sin\bar{\theta}) - \mathbf{k}_{\text{FS}}(\theta_3)}} \,. \quad (S29b)$$

Here $\partial_T f(x)$ is the temperature derivative of the Fermi function, reading $\partial_T f(x) = \frac{x}{(2T \cosh(\frac{x}{2T}))^2}$. The function $r_{\max}(\theta)$ represents the distance of the hexagonal boundary (that is, the hexagon whose vertices are the $K$ and $K'$ points) of the Brillouin zone from the $\Gamma$-point, and it is given by

$$r_{\max}(\theta) = \frac{2\pi}{\sqrt{3}} \frac{1}{\cos\left(\text{mod}\left(\theta + \frac{\pi}{6}, \frac{\pi}{3}\right) - \frac{\pi}{6}\right)} \,. \quad (S30)$$

The initial conditions for the vertices are those given in Eq. (S26). The one-dimensional integrals in Eqs. (S29) are performed discretizing the interval $[0, r_{\max}(\theta)]$ into up to $5 \times 10^4$ steps, depending on the temperature.

Our implementation resolves the Fermi-surface well and treats different types of correlations on equal footing. In this sense, it provides an unbiased of all Fermi surface instabilities. The employed approximations are valid as long as dominant interaction corrections come from low-energy states around the Fermi surface. The small critical temperatures we find (see below) support this scenario.

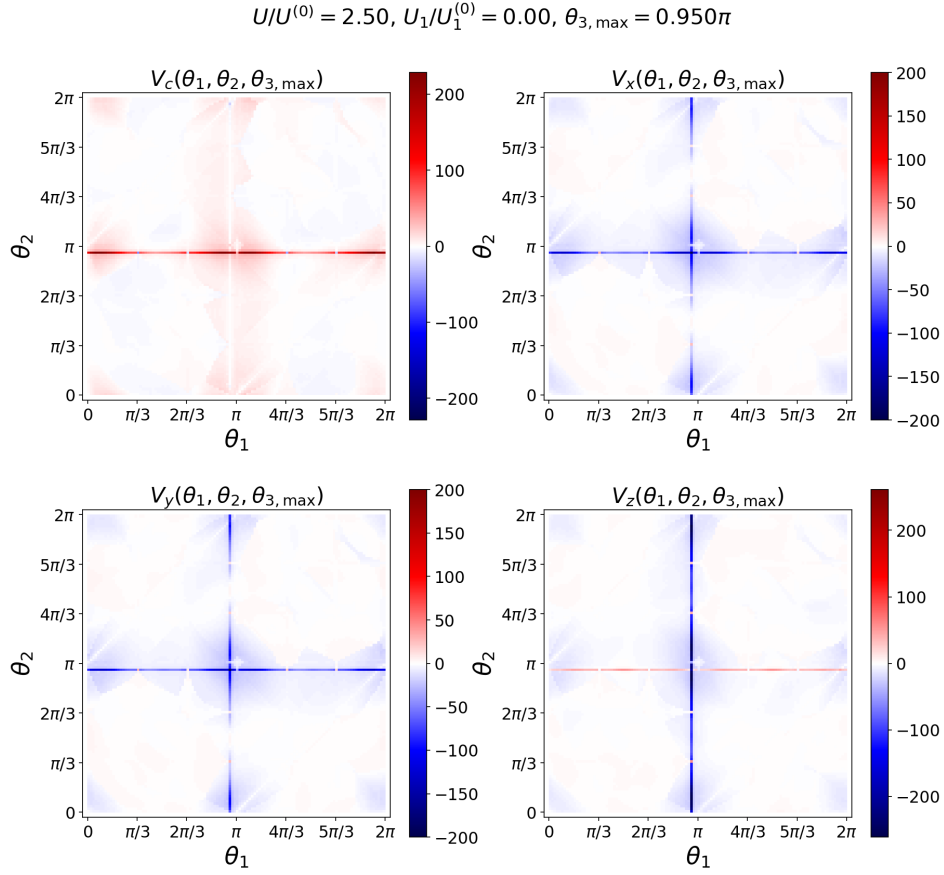

Figure S12. Typical vertex structure of a $q = 0$ instability. Here, $V_z(\theta_1, \theta_2, \theta_{3,\max})$ displays a strong feature along the $\theta_1 = \theta_{3,\max}$ line. This fact, together with the fact that $V_z(\theta_{3,\max}, \theta_2, \theta_{3,\max})$ is approximately a constant, indicates that a ferromagnetic instability is occurring, in which the spins will align along the $z$-direction. A subdominant feature is present also in $V_x$ and $V_y$.

### A. Instability analysis

A Fermi-surface instability occurs when the renormalised interaction diverges. If this happens, one can infer the corresponding type of instability from the form of the singularity. We stop the flow when one of $V_{\uparrow\uparrow}$ or $V_{\uparrow\downarrow}$ exceeds the threshold value of 200 eV. We interpret the temperature scale $T_s$ at which this happens as a mean-field-like critical temperature, below which an order parameters begins to form. To infer the specific form of the order parameter, we consider the following quantities

$$V_c(\theta_1, \theta_2, \theta_3) = V_{\uparrow\uparrow}(\theta_1, \theta_2, \theta_3) + V_{\uparrow\downarrow}(\theta_1, \theta_2, \theta_3), \tag{S31a}$$

$$V_x(\theta_1, \theta_2, \theta_3) = V_y(\theta_1, \theta_2, \theta_3) = -V_{\uparrow\downarrow}(\theta_2, \theta_1, \theta_3), \tag{S31b}$$

$$V_z(\theta_1, \theta_2, \theta_3) = V_{\uparrow\uparrow}(\theta_1, \theta_2, \theta_3) - V_{\uparrow\downarrow}(\theta_1, \theta_2, \theta_3). \tag{S31c}$$

If the largest value of $V_r(\theta_1, \theta_2, \theta_3)$ occurs for $r = c$, we label the instability as a charge instability. Conversely, if it occurs for $r = x, y, z$, we label it as a spin instability along the $x$, $y$ or $z$ direction.

We then fix $\theta_3 = \theta_{3,\max}$, with $\theta_{3,\max}$ the value of $\theta_3$ that maximizes

$$\max_{\theta_1, \theta_2, \theta_3} \left\{ \max \left[ V_{\uparrow\uparrow}(\theta_1, \theta_2, \theta_3), V_{\uparrow\downarrow}(\theta_1, \theta_2, \theta_3) \right] \right\}. \tag{S32}$$

We analyze the quantities $V_r(\theta_1, \theta_2, \theta_3 = \theta_{3,\max})$ as functions of $\theta_1$, $\theta_2$, and $r = c, x, y, z$. If $V_r(\theta_1, \theta_2, \theta_{3,\max})$ takes large values for a $\theta_{1,\max}$ such that $\theta_{1,\max} = \theta_{3,\max}$ independently of $\theta_2$, the instability is a zero momentum transfer (or $q = 0$) instability. Note that the vertex functions defined above are not antisymmetric upon exchanging $\theta_1 \leftrightarrow \theta_2$. Moreover, the information contained in them is redundant, as the full vertex function can be described in terms of only two functions [see Eq. (S16)]. For this reason, we can ignore the fact that some of the $V_{c,x,y,z}$ take large values

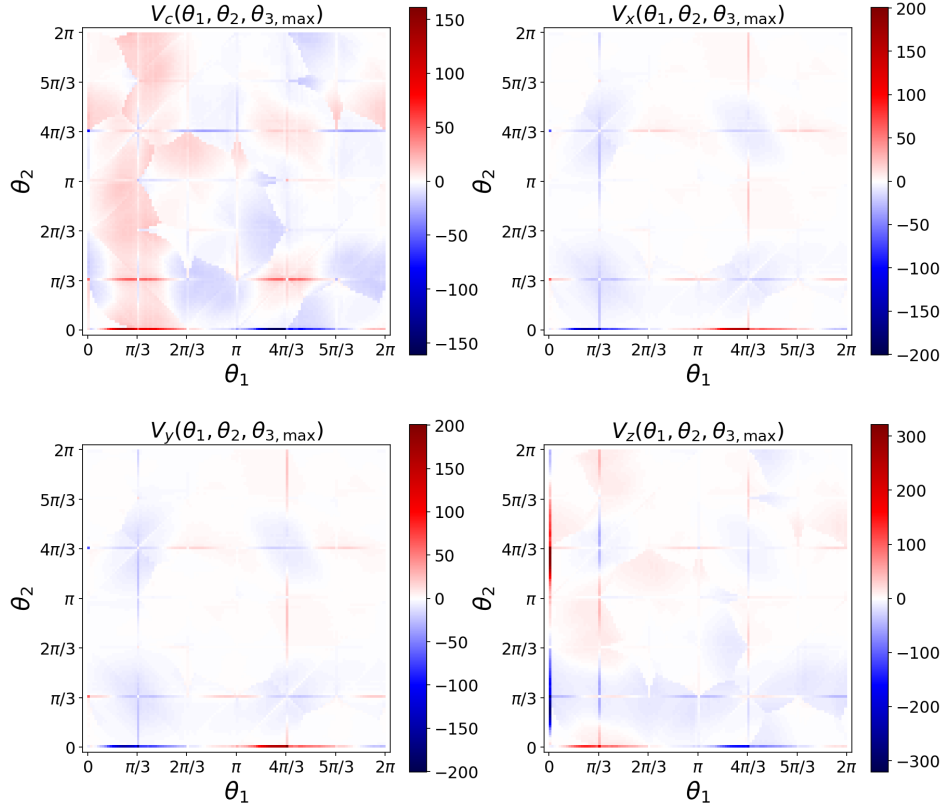

Figure S13. Typical vertex structure of a $q = M$ instability. Here, $V_z(\theta_1, \theta_2, \theta_{3,\max})$ displays a strong feature along the $\theta_1 = 0$ line. Since here $\theta_{3,\max} = \pi/3$, we can infer that the vertex structure here presented indicates an SDW instability. The angular dependence of $V_z(0, \theta_2, \theta_{3,\max})$ will give the momentum structure of the order parameter across the Fermi surface.

for $\theta_2 = \theta_{3,\max}$ as this means that another $V$ will take large values for $\theta_1 = \theta_{3,\max}$. Fixing also $\theta_1 = \theta_{1,\max}$ provides information on the angular dependence of the order parameter. An example of such instability is displayed in Fig. S12.

A different case occurs when $\theta_{3,\max}$ is located at a Van Hove singularity ($n\pi/6$, with $n = 0, 1, 2, \ldots 5$) and the vertices $V_r(\theta_1, \theta_2, \theta_3 = \theta_{3,\max})$ take large values for a value of $\theta_1$ that is located at another Van Hove singularity. In this case, the flow is unstable upon the formation of an order parameter that breaks translational invariance. This order parameter is characterized by at least one of the three wave-vectors that map an $M$-point onto another: $q_1 = (0, 2/\sqrt{3})\pi$, and $(1, \pm 1/\sqrt{3})\pi$. Note that these precisely coincide with the coordinates of the $M$-points themselves, which is why we name this instability a $q = M$ instability. An example of such a behavior of the vertex is displayed in Fig. S13.

Another instability can occur that differs from the ones described above. In some cases $V_r(\theta_1, \theta_2, \theta_{3,max})$ has a "resonance line" for $\theta_2 = \pi + \theta_1$ (modulo $2\pi$), as displayed in Fig. S14. In this case, a superconducting instability occurs, as the instability is at $\mathbf{k}_2 = -\mathbf{k}_1$.

In some cases when is not possible to distinguish between a $q = 0$, a $q = M$ and a superconducting instability, we label the state emerging below $T_s$ as a *coexistence* state of all three kind of orders. A typical vertex structure for this case is shown in Fig. S15.

## SV. PHASE DIAGRAM

In this section, we present the phase diagram, as obtained from $N$-patch fRG, as function of the ratios $U/U^{(0)}$ and $U_1/U_1^{(0)}$, where $U$ and $U_1$ are the onsite Hubbard interaction and nearest-neighbor interaction, respectively, and $U^{(0)} = 5.72$ eV, $U_1^{(0)} = 2.30$ eV are their computed values from cRPA. We kept the $U'_{\sigma,\sigma'}$ and $J_\sigma$ parameters fixed to their cRPA values (see Tab. SI).

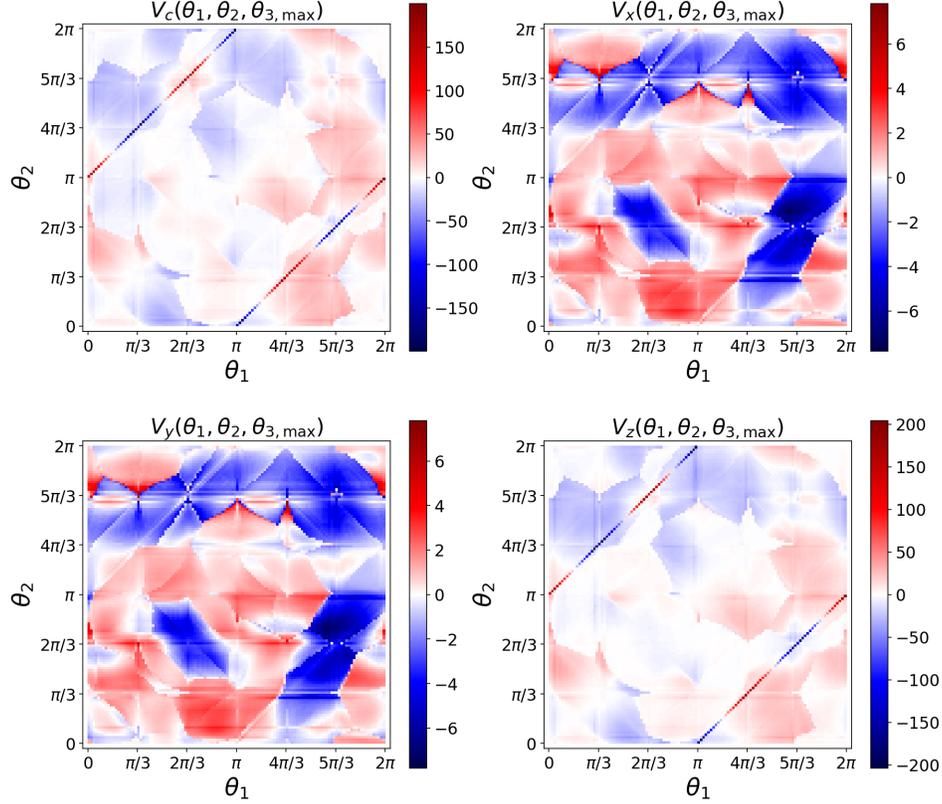

Figure S14. Typical vertex structure of a superconducting instability. Here, $V_z(\theta_1, \theta_2, \theta_{3,\max})$ displays a strong feature along the $\theta_2 = \pi + \theta_1$ (modulo $2\pi$) line. The angular dependence of $V_r(\theta_1, \pi + \theta_1, \theta_{3,\max})$ will give the internal momentum structure of the order parameter.

In Fig. S16, we show the obtained phase diagram. For small values of $U$ and $U_1$, we find no instability down to the lowest numerically accessible temperature, $T \sim 1.10~\mu$eV $= 0.012$ K. This is indicated in Fig. S16 by the black crosses (corresponding to the computed points in the phase diagram) with white background. Note that even for $U = U_1 = 0$ we are dealing with an interacting problem due to the presence of nonzero $U'$ and $J$.

For $U_1 = 0$ and $U \geq U^{(0)}$, we find that the system is unstable upon the formation of ferrimagnetic order along the $z$-axis, that is, the axis along which the *antiferromagnetic* order parameter is aligned. This ordering spontaneously breaks the discrete symmetry $T_z \otimes \mathcal{T}$ of the antiferromagnet, giving a magnetic phase in which the spins in neighboring layers point in opposite directions but they additionally have different *magnitudes*. Moreover, the two layers will have different average charge densities. We indicated this phase in Fig. S16 by black crosses with dark blue background.

For $U_1 = 0.5\, U_1^{(0)}$ and $U^{(0)} \leq U \leq 1.5\, U^{(0)}$, we find a *spin-Pomeranchuk* instability. This phase, similar to the aforementioned ferromagnetic one, breaks $T_z \otimes \mathcal{T}$ but such that the momentum averaged order parameter $\int_{\mathbf{k}} \langle a_{\mathbf{k}}^\dagger \sigma^z a_{\mathbf{k}} \rangle = 0$, with $\sigma^z = \left(\begin{smallmatrix} 1 & 0 \\ 0 & -1 \end{smallmatrix}\right)$ the third Pauli matrix. The two most negative eigenvalues of the matrix $V_z^{q=0}(\theta_1, \theta_2) \equiv V_z(\theta_1, \theta_2, \theta_1)$, with their eigenvectors (displayed in Fig. S17) corresponding to the Fermi surface projections of the $d_{xy}$ and $d_{x^2-y^2}$ form factors. Below $T_s$, the system will develop an order parameter that is a linear combination of the two, thereby spontaneously breaking the lattice $D_{6h}$ symmetry. We note that in this regime of the phase diagram, the divergence of the vertex is strongly centered around Van Hove points. If we only take these peaks at Van Hove points into account without the subleading wave vector dependence, $d$-wave charge and spin-Pomeranchuk are of the same order. The Pomeranchuk phase is indicated by black crosses with orange background in Fig. S16. Fourier transforming the two basis functions displayed in Fig. S17, we obtain Fig. S18, indicating that the spin-Pomeranchuk phase is essentially an *inter-sublattice* SDW. The low-energy theory for the spin-Pomeranchuk phase involves a scalar complex order parameter $\phi \sim \int_{\mathbf{k}} (d_{\mathbf{k}}^{x^2-y^2} + i d_{\mathbf{k}}^{xy}) a_{\mathbf{k}}^\dagger \sigma^z a_{\mathbf{k}}$, with $d_{\mathbf{k}}^{x^2-y^2} = (2/\sqrt{3})[\cos k_x - \cos(k_x/2)\cos(\sqrt{3}k_y/2)]$ and $d_{\mathbf{k}}^{xy} = 2\sin(k_x/2)\sin(\sqrt{3}k_y/2)$. Under a lattice rotation around the center of a hexagonal plaquette by an angle of

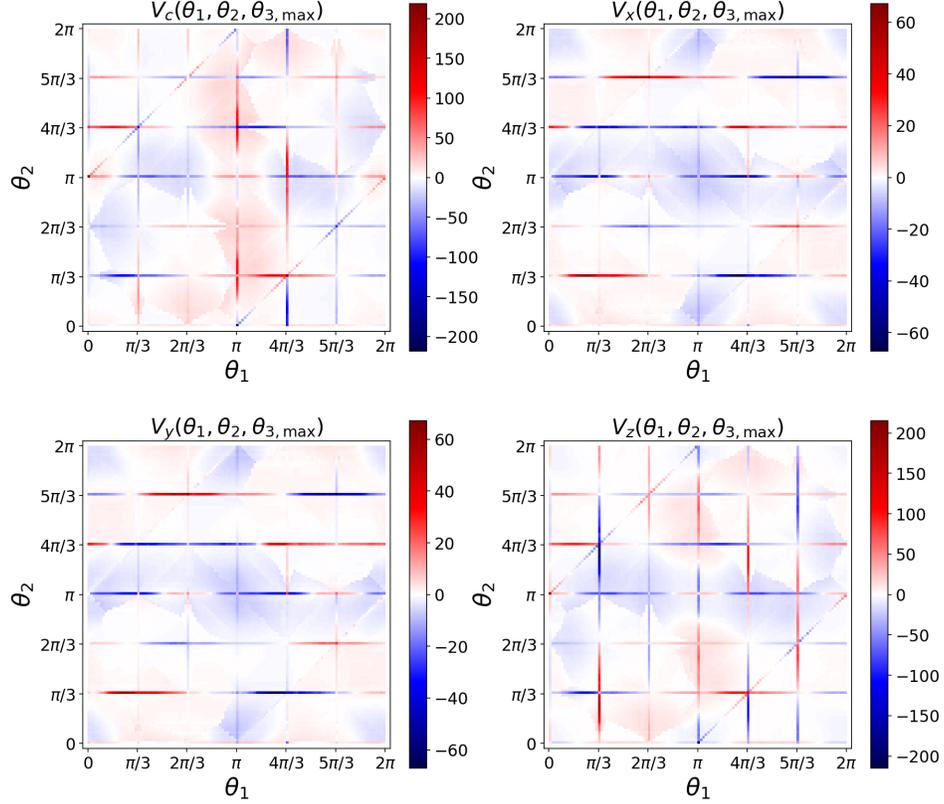

Figure S15. Vertex structure in a case in which it is impossible to distinguish between a $q = 0$, a $q = M$, and a superconducting instability. Note that $V_c$ and $V_z$ display vertical features not only at $\theta_1 = \theta_{3,\max}$ but also at $\theta_1 = \theta_{3,\max} + \frac{2n\pi}{3}$ with $n = -2, -1, 1, 2$. Additionally, similarly to Fig. S14, we have a diagonal feature in $V_z$, which is a signature of a concomitant superconducting instability.

$n\pi/6$, $\phi$ transforms as $e^{i\frac{2n\pi}{6}}$, where the factor of two in the phase factor comes from the fact that we are dealing with an order parameter with angular momentum 2. Upon applying a $T_z \otimes \mathcal{T}$ transformation, $\phi$ gets a minus sign. Putting these symmetry properties together, we can write down the Landau free energy of our spin-Pomeranchuk order parameter:

$$\mathcal{F}[\phi] = r|\phi|^2 + u_4 \left(|\phi|^2\right)^2 + u_6 \left(|\phi|^2\right)^3 + v \left[\phi^6 + (\phi^*)^6\right] . \tag{S33}$$

Note that a term of the form $\phi^3 + (\phi^*)^3$ would be allowed if we were dealing with a *charge*-Pomeranchuk (or nematic) instability, but it is forbidden by the $T_z \otimes \mathcal{T}$ symmetry in this case. A negative value of $v$ implies that the order parameter will take the form $\phi \propto e^{i\frac{n\pi}{3}}$, corresponding to the left panel of Fig. S18, modulo a lattice rotation or a flip of all spins. Conversely, $v > 0$ gives $\phi \propto ie^{i\frac{n\pi}{3}}$, which corresponds to the right panel of Fig. S18 or to a symmetry equivalent state. The sign of $v$ can be estimated by coupling free electrons with dispersion $\epsilon_{\mathbf{k}}$ in Eq. (S9) to the $\phi$ order parameter, integrating them out and expanding the resulting effective action up to sixth order in $\phi$, which is beyond the scope of the current study. The spin-Pomeramchuk state, additionally, will induce a charge density wave, both *intra* and *inter*layer. If $\vec{S}_i$ is the onsite spin density, given by the antiferromagnetic contribution $M(-1)^{i_z}\hat{e}_z$ and the spin-Pomerachuk term $S_{\text{pom}}f_{i_x,i_y}\hat{e}_z$ (with $f_{i_x,i_y}$ being one of the two patterns displayed in Fig. S18 and $M$ and $S_{\text{pom}}$ real numbers quantifying the strength of the two orders), the modulation of the charge density will depend on the spatial coordinate as $\delta\rho_i \propto |\vec{S}_i|^2 - M^2 \simeq 2(-1)^{i_z} M S_{\text{pom}} f_{i_x,i_y}$, assuming $M \gg S_{\text{pom}}$.

For $U_1 = 0.5\, U_1^{(0)}$ and $U \geq 1.75\, U^{(0)}$, we find another ferromagnetic phase, in which the order parameter lies in the $xy$ plane, that is, perpendicular to the axis along which the antiferromagnetic order is aligned. This spontaneous ordering, similarly to the previously discussed ones, breaks the discrete symmetry $T_z \otimes \mathcal{T}$, as well as the residual U(1) symmetry given by rotations around the $z$-axis. In this phase, the antiferromagnetically ordered spins will get a

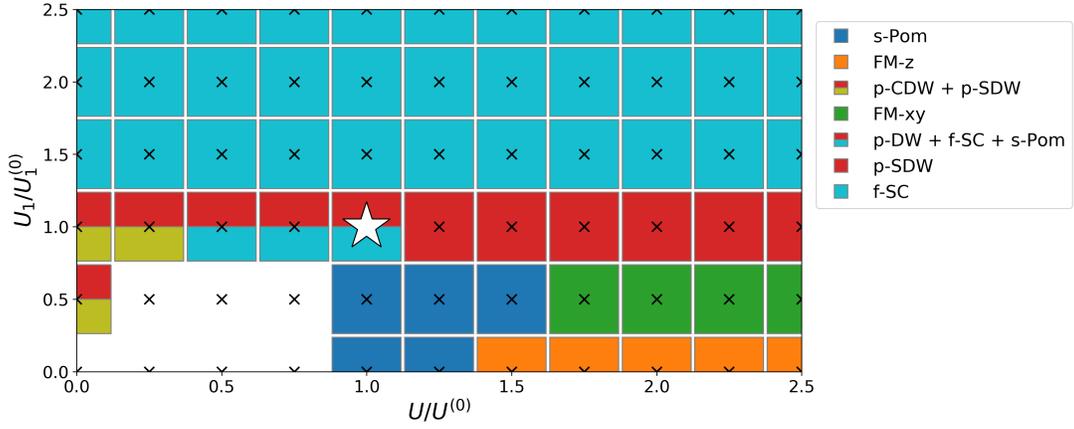

Figure S16. Phase diagram of FeGe as a function of the ratios of the onsite Hubbard $U$ and nearest neighbor interaction $U_1$ with their values computed from cRPA. The abbreviations have the following meanings: FM-z is ferromagnetism along the $z$-axis, p-CDW+p-SDW is a phase with $p$-wave charge and spin density wave order, s-Pom stands for spin Pomeranchuk, FM-xy corresponds to ferromagnetism in the $xy$ plane, p-DW+f-SC+s-Pom is a phase in which $p$-CDW, $p$-SDW, $f$-wave superconductivity and a spin Pomeranchuk phase likely coexist, and f-SC stands for $f$-wave superconductivity. The white star marks the point in the phase diagram corresponding to the cRPA values of the interactions.

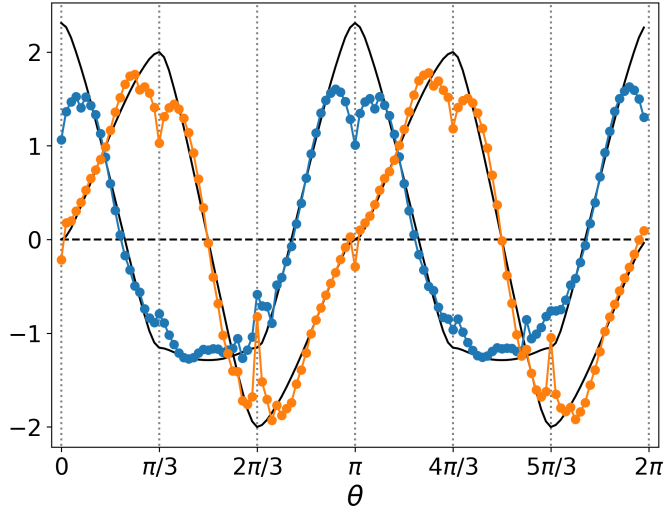

Figure S17. Leading eigenvectors of the matrix $V_z^{q=0}(\theta_1, \theta_2)$ (see text) in the spin-Pomeranchuk phase, obtained for $U = 1.5\,U^{(0)}$ and $U_1 = 0.5\,U_1^{(0)}$. The black lines represent the functions $d_{\mathbf{k}}^{x^2-y^2} = (2/\sqrt{3})[\cos k_x - \cos(k_x/2)\cos(\sqrt{3}k_y/2)]$ and $d_{\mathbf{k}}^{xy} = 2\sin(k_x/2)\sin(\sqrt{3}k_y/2)$.

canting, pointing in the same direction in both layers. Since the amplitude of the spins is the same in each layer, no modulation of the charge density will be induced.

For $U_1 \geq 1.5 U_1^{(0)}$ and independent of the value of $U$, we find a superconducting instability. The internal structure of the order parameter can be in inferred by analyzing the eigenvectors of the matrices $V_{\uparrow\uparrow}(\theta_1, \pi + \theta_1, \theta_3)$ and $V_{\uparrow\downarrow}(\theta_1, \pi + \theta_1, \theta_3)$. We find that the largest negative eigenvalue of $V_{\uparrow\uparrow}$ is much larger in absolute value than the one of $V_{\uparrow\downarrow}$, indicating a *spin-triplet* superconducting instability, in which equal spins are paired. The eigenvector corresponding to the largest negative eigenvalue of $V_{\uparrow\uparrow}(\theta_1, \pi + \theta_1, \theta_3)$ is shown in Fig. S19. We notice that it possesses the same symmetry as the higher-order $f$-wave form factor $f_{\mathbf{k}}^f \propto \sin(\sqrt{3}k_y) - 2\cos(3k_x/2)\sin(\sqrt{3}k_y/2)$, whose Fourier transform $f_{ij}^f$ is nonzero only if $i$ and $j$ are next-to-nearest-neighboring sites (see Fig. 3(f) of the main text). Higher harmonics (involving further distant sites) also seem to play an important role (see Fig. S19). Note that a superconducting instability in the lowest order $f$-wave channel $f_{\mathbf{k}}^{f_o} \propto \sin(k_x) - 2\sin(k_x/2)\cos(\sqrt{3}k_y/2)$ (coupling nearest-neighboring sites) is not allowed due to the repulsive nearest-neighbor interaction. Within our approach, different combinations

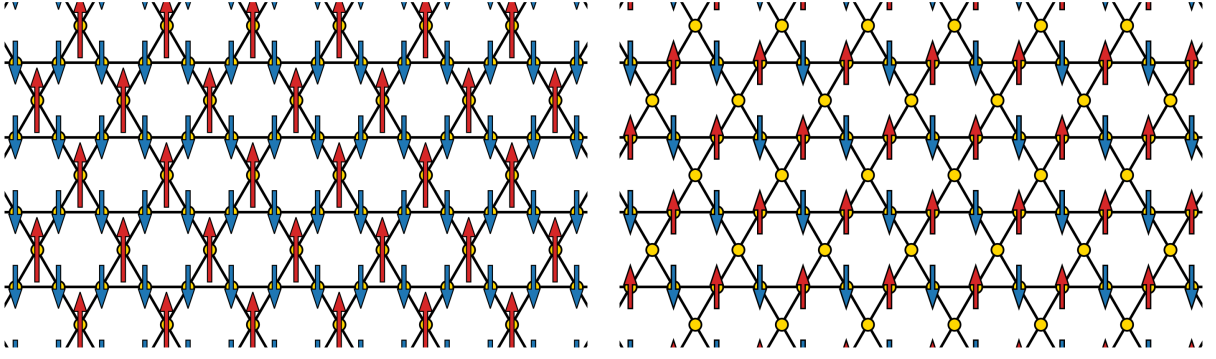

Figure S18. Fourier transform of the two basis functions in Fig. S17. In the spin-Pomeranchuk phase the system realizes a spin-$z$ ordering which is one of the two patterns shown here, possibly rotated by an angle multiple of $\frac{2\pi}{3}$ around the center of a hexagonal plaquette. Longer arrows represent a higher onsite magnetization. Note that we are subtracting the background magnetization, due to the underlying Néel state.

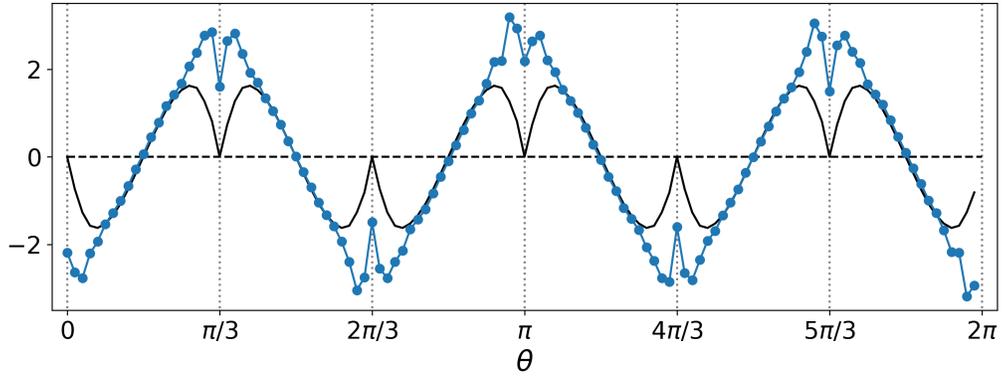

Figure S19. Leading eigenvector of the matrix $V_{\uparrow\uparrow}(\theta_1, \pi + \theta_1, \theta_3)$ in the superconducting phase, obtained for $U = 2\,U^{(0)}$ and $U_1 = 1.5\,U_1^{(0)}$. The solid black line represents the function $f_{\mathbf{k}}^f \propto \sin(\sqrt{3}k_y) - 2\sin(3k_x/2)\sin(\sqrt{3}k_y/2)$.

of the triplet components $(\cos\theta\sigma_x + e^{i\varphi}\sin\theta\sigma_y)i\sigma_y$ are degenerate. Below the critical temperature, the combination with the lowest energy is realized. This can spontaneoulsy break the symmetry between the layers in a non-unitary pairing state with $\theta = \pi/4, \varphi = \pi/2$. Note that time-reversal is already broken through the antiferromagnetic order between the layers. To understand the pairing mechanism better, we analyze the bare pairing vertex and find that there is a small attractive $f$-wave component after projection into the $\alpha$ band which induces the pairing instability, which is then further enhanced by fluctuations in the particle-hole channels. The superconducting phase is indicated by black crosses on a cyan background in Fig. S16.

For $U_1 = U_1^{(0)}$ and independent of the value of $U$, we find a $q = M$ instability. For smaller values of $U$, this instability appears both in $V_c$ and $V_z$, while for larger values of $U$, it only appears in $V_z$. To infer the internal structure order parameter, we consider the eigenvector of the matrices

$$V_r^{(n)}(\theta_1, \theta_2) \equiv V_r(\theta_1, \theta_2, \text{proj}_{\text{FS}}[\mathbf{k}_{\text{FS}}(\theta_1) + \mathbf{Q}_n]),  \quad (S34)$$

with $n = 1, 2, 3$ and $\mathbf{Q}_1 = (0, (2\pi)/\sqrt{3})$, $\mathbf{Q}_2 = (\pi, -\pi/\sqrt{3})$, $\mathbf{Q}_3 = (-\pi, -\pi/\sqrt{3})$. In order to ensure that $\mathbf{k}(\theta_1) + \mathbf{Q}_n$ is not excessively far from the Fermi surface, we restrict $\theta_1$ and $\theta_2$ to

$$\theta_1, \theta_2 \in \left[\frac{\pi}{3}, \frac{2\pi}{3}\right] \cup \left[\frac{4\pi}{3}, \frac{5\pi}{3}\right] \quad \text{for } n = 1, \quad (S35a)$$

$$\theta_1, \theta_2 \in \left[0, \frac{\pi}{3}\right] \cup \left[\pi, \frac{4\pi}{3}\right] \quad \text{for } n = 2, \quad (S35b)$$

$$\theta_1, \theta_2 \in \left[\frac{2\pi}{3}, \pi\right] \cup \left[\frac{5\pi}{3}, 2\pi\right] \quad \text{for } n = 3. \quad (S35c)$$

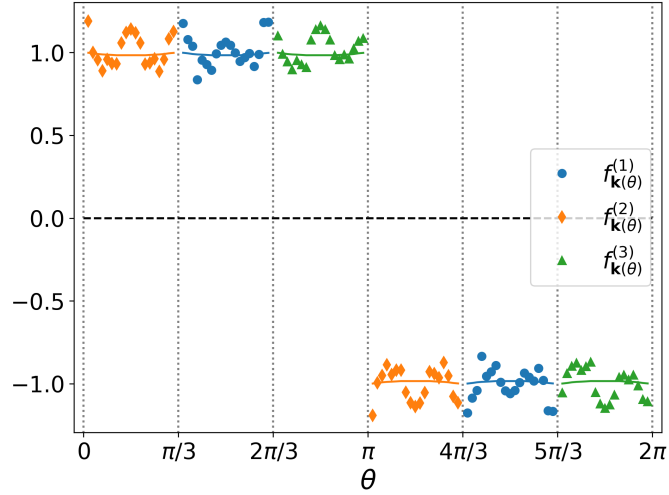

Figure S20. Eigenvectors $f^{(n)}_{\mathbf{k}(\theta)}$ of the matrices $V_z^{(n)}(\theta_1, \theta_2)$ corresponding to the most negative eigenvalue. The solid lines represent the fitting functions (S36). The parameters are $U/U^{(0)} = 0$, $U_1/U_1^{(0)} = 1$.

In Fig. S20, we show the eigenvectors $f^{(n)}_{\mathbf{k}(\theta)}$ of the matrices $V_z^{(n)}(\theta_1, \theta_2)$ corresponding to the most negative eigenvalue. We see that the functions that qualitatively best reproduce the behavior of the three eigenvectors are

$$f^{(1)}_{\mathbf{k}} = -\sin\left(\frac{k_x}{2}\right), \tag{S36a}$$

$$f^{(2)}_{\mathbf{k}} = \sin\left(\frac{k_x + \sqrt{3}k_y}{4}\right), \tag{S36b}$$

$$f^{(3)}_{\mathbf{k}} = \sin\left(\frac{k_x - \sqrt{3}k_y}{4}\right). \tag{S36c}$$

We indicated them in Fig. S20 with solid lines.

The ordered phase emerging below $T_s$ in this case is characterized by three order parameters,

$$\Phi_n = \int_{\mathbf{k}\in\mathrm{FS}} f^{(n)}_{\mathbf{k}} \langle a^\dagger_{\mathbf{k}}\, \eta\, a_{\mathbf{k}+\mathbf{Q}_n}\rangle, \tag{S37}$$

with $\eta = \mathbb{1}$ or $\sigma^z$, and $f^{(n)}_{\mathbf{k}}$ defined in Eqs. (S36). Because of the specific form of the form factors entering Eq. (S37), we name this phase a $p$-wave charge density wave ($p$-CDW) or spin density wave ($p$-SDW).

This state breaks the $T_z \otimes \mathcal{T}$ symmetry, as the order parameter takes the same form in both layers, but not on average, that is, $\int_{\mathbf{k}\in\mathrm{FS}} \langle a^\dagger_{\mathbf{k}} \sigma^z a_{\mathbf{k}+\mathbf{Q}_n}\rangle = 0$ for $n = 1, 2, 3$, as well as translational symmetry. Inspecting Eq. (S37), we see that it describes charge ($\eta = \mathbb{1}$) or spin-$z$ ($\eta = \sigma^z$) bond density waves. In the ordered state, one or more of the order parameters in Eq. (S37) can condense. Unless all three order parameters condense with the same absolute value, the lattice $D_{6h}$ symmetry will be broken, too. In Fig. S21 we compare the charge and spin bond density patterns in the case in which only $\Phi_1$ is finite (left panel) and in the case in which $\Phi_1$, $\Phi_2$, and $\Phi_3$ are all finite and possess the same absolute value (right panel). In the first case, we find that the system exhibits an alternation of enhanced and suppressed bond densities along the $(1, 0)$-direction. Moreover, the unit cell (gray dashed line in Fig. S21) doubles in size and contains 6 lattice sites.

If all order parameters condense, $D_{6h}$ symmetry remains preserved, and "rings" of enhanced bond density appear around every fourth triangular and hexagonal plaquette of the kagome lattice (see Fig. S21). Around a "star of David" path, instead, we find suppressed bond densities (Fig. S21). The unit cell (gray dashed lines in Fig. S21) size increases by a factor of 4 and contains 12 kagome sites.

In order to determine whether a the phase depicted in the left or in the right panel of Fig. S21 is energetically favored, we resort to a Landau analysis. A mean-field potential for the three order parameters in Eq. (S37) takes the

form

$$\mathcal{F}[\Phi_1, \Phi_2, \Phi_3] = r \sum_{n=1}^{3} \Phi_n^2 + u \left(\sum_{n=1}^{3} \Phi_n^2\right)^2 + \lambda \left(\Phi_1^2 \Phi_2^2 + \Phi_2^2 \Phi_3^2 + \Phi_3^2 \Phi_1^2\right), \quad \text{(S38)}$$

where $C_6$ rotations act as permutations of the sequence $(\Phi_1, \Phi_2, \Phi_3)$ and the $T_z \otimes \mathcal{T}$ symmetry changes the sign of all three $\Phi_n$. One can convince himself that the form above of the Landau potential contains all symmetry allowed terms up to quartic order. Despite them having zero momentum, third-order terms of the form $\Phi_1 \Phi_2 \Phi_3$ are not allowed because they are odd under the action of $T_z \otimes \mathcal{T}$ symmetry operations. If the interaction $\lambda$ between different order parameters is negative, then the energetically favored phase will have all $\Phi_n$ condensed and with the same amplitude. Vice-versa, if it is positive only one $\Phi_n$ will be nonzero. $\lambda$ can be calculated with the sole knowledge of the band dispersion $\epsilon_\mathbf{k}$ by integrating out the fermions and expanding in $\Phi_n$. A low energy theory of Fermions coupled to the (classical) order parameters $\Phi_n$ takes the form

$$\mathcal{S}[\psi, \overline{\psi}, \Phi_n] = T \sum_\omega \sum_\sigma \int_\mathbf{k} \overline{\psi}_{\mathbf{k},\omega,\sigma}(i\omega - \epsilon_\mathbf{k}) \psi_{\mathbf{k},\omega,\sigma} + \int_\mathbf{k} T \sum_\omega \sum_{\substack{n=1,2,3 \\ \sigma}} \Phi_n f_\mathbf{k}^{(n)} \overline{\psi}_{\mathbf{k},\omega,\sigma} \psi_{\mathbf{k}+\mathbf{Q}_n,\omega,\sigma} + \frac{1}{g_0} \sum_{n=1,2,3} \Phi_n^2, \quad \text{(S39)}$$

where $\psi$ and $\overline{\psi}$ are Grassmann variables, and $g_0$ is an effective coupling that can be evaluated microscopically, even though we will not do it here. Integrating out the fermions, one gets the effective action

$$\mathcal{S}_{\text{eff}}[\Phi_n] = \text{Trln}\left[(i\omega - \epsilon_\mathbf{k})\delta_{\omega,\omega'}\delta_{\mathbf{k},\mathbf{k}'}\delta_{\sigma\sigma'} - \sum_n \Phi_n f_\mathbf{k}^{(n)} \delta_{\omega,\omega'}\delta_{\mathbf{k}'=\mathbf{k}+\mathbf{Q}_n}\delta_{\sigma\sigma'}\right], \quad \text{(S40)}$$

where the argument of the logarithm has to be intended as a matrix in momentum, frequency, and spin variables, and a trace over these variables is taken at the end. Expanding the effective action up to quartic order in $\Phi_n$ and comparing term by term with Eq. (S38), we can obtain microscopic expressions for the coefficients of the Landau potential. For $\lambda$, we obtain

$$\lambda = \int_{\mathbf{k} \in \text{FS}} \left[ 2 f_\mathbf{k}^{(1)} f_{\mathbf{k}+\mathbf{Q}_1}^{(1)} f_\mathbf{k}^{(2)} f_{\mathbf{k}+\mathbf{Q}_2}^{(2)} F(\epsilon_\mathbf{k}, \epsilon_{\mathbf{k}+\mathbf{Q}_1}, \epsilon_\mathbf{k}, \epsilon_{\mathbf{k}+\mathbf{Q}_2}) \right. \\
+ f_\mathbf{k}^{(1)} f_{\mathbf{k}+\mathbf{Q}_1}^{(2)} f_{\mathbf{k}+\mathbf{Q}_3}^{(1)} f_{\mathbf{k}+\mathbf{Q}_2}^{(2)} F(\epsilon_\mathbf{k}, \epsilon_{\mathbf{k}+\mathbf{Q}_1}, \epsilon_{\mathbf{k}+\mathbf{Q}_3}, \epsilon_{\mathbf{k}+\mathbf{Q}_2}) \\
\left. - \left[f_\mathbf{k}^{(1)}\right]^2 \left[f_{\mathbf{k}+\mathbf{Q}_1}^{(1)}\right]^2 F(\epsilon_\mathbf{k}, \epsilon_{\mathbf{k}+\mathbf{Q}_1}, \epsilon_\mathbf{k}, \epsilon_{\mathbf{k}+\mathbf{Q}_1}) \right], \quad \text{(S41)}$$

with

$$F(\epsilon_1, \epsilon_2, \epsilon_3, \epsilon_4) = \sum_{i=1}^{4} \frac{f(\epsilon_i)}{\Pi_{j \neq i}(\epsilon_i - \epsilon_j)}, \quad \text{(S42)}$$

with $f(x)$ the Fermi function. Note that, for the sake of consistency with the patch-fRG calculation, we have restricted the above momentum integral to the Fermi surface. Our calculations show that $\lambda < 0$, meaning that the state with all $\Phi_n$ condensed is energetically favored. Independent of whether or not all 3 order parameter condense, the resulting state breaks $T_z \otimes \mathcal{T}$ and the intralayer translational symmetry. If only one of the $\Phi_n$ condenses, the intralayer point group symmetry $D_{6h}$ is additionally broken.

In Fig. S21, we indicate with black crosses with red and gold background the points in the phase diagram in which instabilities towards both a $p$-CDW and a $p$-SDW (along the $z$-direction) will be formed. At the points indicated by black crosses on a red background, the $p$-SDW instability is clearly leading. We note again that spin-$z$ and charge instabilities do not decouple, i.e., even if one is the leading instability here, if one of them develops, the other must too. Finally, in the regime indicated by black crosses on a red and cyan background, the vertices exhibit sharp features typical of both a $p$-CDW/$p$-SDW phase, of the $f$-wave superconducting one, and of the Pomeranchuk one. This is the regime, where the ab initio values place the interaction parameters of FeGe. Interestingly, in contrast to other regions of the phase diagram, we cannot distinguish a clear winner of the competing orders here. Instead we find a strongly correlated regime where all types of correlations grow large.

### A. Critical temperature

In Fig. S22, we show the value of the fRG critical temperature $T_c$ as a function of the interactions $U$ and $U_1$. We observe that $T_c$ depends strongly on $U_1$, displaying a mild dependence on $U$ only for $U_1 \lesssim 1.5 \, U_1^{(0)}$. We note that we

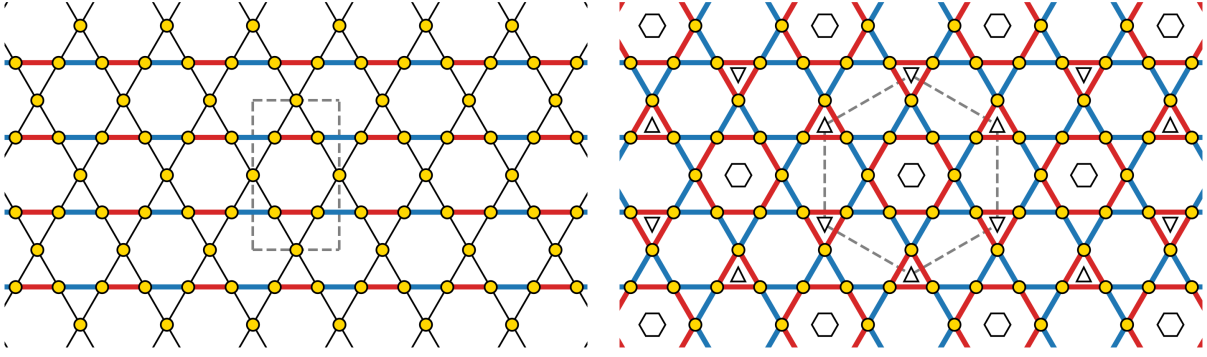

Figure S21. Spin-$z$ and charge bond density patterns on the Fe kagome lattice for the $p$-CDW/$p$-SDW phase in which only $O^{(1)}$ condenses (left panel), and in which all three order parameters in Eq. (S37) condense with the same amplitude (right panel). Red (blue) bonds indicate an enhanced (reduced) hopping strength along that bond. In both panels the gray dashed lines indicate the enlarged unit cells.

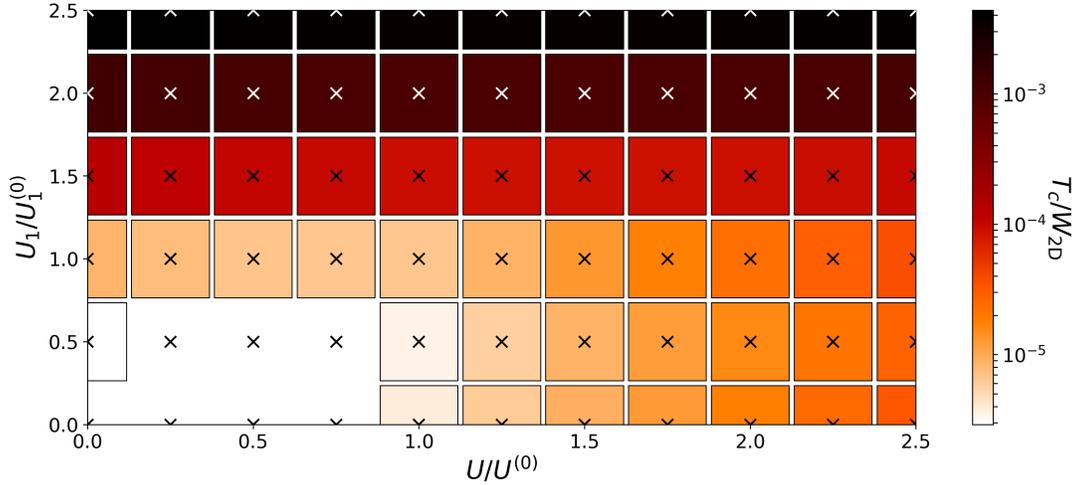

Figure S22. fRG critical temperature $T_c$ as a function of the onsite ($U$) and nearest-neighbor ($U_1$) interactions.

always obtain very small values of the critical temperature compared to the two-dimensional bandwidth $W_{2D} \sim 450$ meV$\sim 5000$ K, which justifies our fRG approach *a posteriori*.



\end{document}